\documentclass[longauth]{aa}

\usepackage{graphicx}
\usepackage{natbib}
\usepackage{scalerel}
\usepackage{comment}

\usepackage[table]{xcolor}

\usepackage{txfonts}
\usepackage[pdfencoding=auto,psdextra]{hyperref}
\hypersetup{
    colorlinks=true,
    linkcolor=blue,
    filecolor=magenta,      
    urlcolor=blue,
    citecolor=blue
}
\makeatletter
\renewcommand*\aa@pageof{, page \thepage{} of \pageref*{LastPage}}
\makeatother

\usepackage[utf8]{inputenc}

\usepackage[switch, modulo]{lineno}

\renewcommand{\linenumbers}{}

\usepackage{euclid}

\begin{document}

    \title{Euclid Quick Data Release (Q1)}
    \subtitle{XLVI. The metallicity of the intracluster light}    

\newcommand{\orcid}[1]{} 			   
\author{Euclid Collaboration: J.~B.~Golden-Marx\orcid{0000-0002-6394-045X}\thanks{\email{icl-group@nottingham.ac.uk}}\inst{\ref{aff1}}
\and N.~A.~Hatch\orcid{0000-0001-5600-0534}\inst{\ref{aff1}}
\and S.~P.~Bamford\orcid{0000-0001-7821-7195}\inst{\ref{aff1}}
\and H.~Gao\orcid{0000-0003-1015-5367}\inst{\ref{aff1}}
\and T.~Kolcu\orcid{0000-0002-3365-9210}\inst{\ref{aff1}}
\and M.~Montes\orcid{0000-0001-7847-0393}\inst{\ref{aff2}}
\and H.~J.~Brown\orcid{0009-0004-7521-8204}\inst{\ref{aff1}}
\and Y.~M.~Bah\'{e}\orcid{0000-0002-3196-5126}\inst{\ref{aff1},\ref{aff3}}
\and H.~Gully\orcid{0000-0001-7616-7278}\inst{\ref{aff1}}
\and G.~Martin\orcid{0000-0003-2939-8668}\inst{\ref{aff1}}
\and M.~Urbano\orcid{0000-0001-5640-0650}\inst{\ref{aff4}}
\and S.~L.~Ahad\orcid{0000-0001-6336-642X}\inst{\ref{aff5},\ref{aff6}}
\and C.~Bellhouse\orcid{0000-0002-6179-8007}\inst{\ref{aff1}}
\and M.~Bolzonella\orcid{0000-0003-3278-4607}\inst{\ref{aff7}}
\and M.~Butler\orcid{0009-0004-9273-4630}\inst{\ref{aff1}}
\and Weiguang~Cui\orcid{0000-0002-2113-4863}\inst{\ref{aff8},\ref{aff9},\ref{aff10}}
\and F.~Durret\orcid{0000-0002-6991-4578}\inst{\ref{aff11}}
\and A.~Ellien\orcid{0000-0002-1038-3370}\inst{\ref{aff12}}
\and A.~H.~Gonzalez\orcid{0000-0002-0933-8601}\inst{\ref{aff13}}
\and M.~Kluge\orcid{0000-0002-9618-2552}\inst{\ref{aff14}}
\and L.~Moscardini\orcid{0000-0002-3473-6716}\inst{\ref{aff15},\ref{aff7},\ref{aff16}}
\and F.~R.~Pearce\orcid{0000-0002-2383-9250}\inst{\ref{aff1}}
\and R.~Ragusa\inst{\ref{aff17}}
\and B.~Altieri\orcid{0000-0003-3936-0284}\inst{\ref{aff18}}
\and S.~Andreon\inst{\ref{aff19}}
\and N.~Auricchio\orcid{0000-0003-4444-8651}\inst{\ref{aff7}}
\and C.~Baccigalupi\orcid{0000-0002-8211-1630}\inst{\ref{aff20},\ref{aff21},\ref{aff22},\ref{aff23}}
\and M.~Baldi\orcid{0000-0003-4145-1943}\inst{\ref{aff24},\ref{aff7},\ref{aff16}}
\and S.~Bardelli\orcid{0000-0002-8900-0298}\inst{\ref{aff7}}
\and P.~Battaglia\orcid{0000-0002-7337-5909}\inst{\ref{aff7}}
\and A.~Biviano\orcid{0000-0002-0857-0732}\inst{\ref{aff21},\ref{aff20}}
\and E.~Branchini\orcid{0000-0002-0808-6908}\inst{\ref{aff25},\ref{aff26},\ref{aff19}}
\and M.~Brescia\orcid{0000-0001-9506-5680}\inst{\ref{aff27},\ref{aff17}}
\and J.~Brinchmann\orcid{0000-0003-4359-8797}\inst{\ref{aff28},\ref{aff29},\ref{aff30}}
\and S.~Camera\orcid{0000-0003-3399-3574}\inst{\ref{aff31},\ref{aff32},\ref{aff33}}
\and G.~Ca\~nas-Herrera\orcid{0000-0003-2796-2149}\inst{\ref{aff34},\ref{aff35}}
\and V.~Capobianco\orcid{0000-0002-3309-7692}\inst{\ref{aff33}}
\and C.~Carbone\orcid{0000-0003-0125-3563}\inst{\ref{aff36}}
\and V.~F.~Cardone\inst{\ref{aff37},\ref{aff38}}
\and J.~Carretero\orcid{0000-0002-3130-0204}\inst{\ref{aff39},\ref{aff40}}
\and S.~Casas\orcid{0000-0002-4751-5138}\inst{\ref{aff41},\ref{aff42}}
\and M.~Castellano\orcid{0000-0001-9875-8263}\inst{\ref{aff37}}
\and G.~Castignani\orcid{0000-0001-6831-0687}\inst{\ref{aff7}}
\and S.~Cavuoti\orcid{0000-0002-3787-4196}\inst{\ref{aff17},\ref{aff43}}
\and K.~C.~Chambers\orcid{0000-0001-6965-7789}\inst{\ref{aff44}}
\and A.~Cimatti\inst{\ref{aff45}}
\and C.~Colodro-Conde\inst{\ref{aff46}}
\and G.~Congedo\orcid{0000-0003-2508-0046}\inst{\ref{aff9}}
\and C.~J.~Conselice\orcid{0000-0003-1949-7638}\inst{\ref{aff47}}
\and L.~Conversi\orcid{0000-0002-6710-8476}\inst{\ref{aff48},\ref{aff18}}
\and Y.~Copin\orcid{0000-0002-5317-7518}\inst{\ref{aff49}}
\and A.~Costille\inst{\ref{aff50}}
\and F.~Courbin\orcid{0000-0003-0758-6510}\inst{\ref{aff51},\ref{aff52},\ref{aff53}}
\and H.~M.~Courtois\orcid{0000-0003-0509-1776}\inst{\ref{aff54}}
\and M.~Cropper\orcid{0000-0003-4571-9468}\inst{\ref{aff55}}
\and J.-C.~Cuillandre\orcid{0000-0002-3263-8645}\inst{\ref{aff56}}
\and A.~Da~Silva\orcid{0000-0002-6385-1609}\inst{\ref{aff57},\ref{aff58}}
\and H.~Degaudenzi\orcid{0000-0002-5887-6799}\inst{\ref{aff59}}
\and G.~De~Lucia\orcid{0000-0002-6220-9104}\inst{\ref{aff21}}
\and H.~Dole\orcid{0000-0002-9767-3839}\inst{\ref{aff60}}
\and F.~Dubath\orcid{0000-0002-6533-2810}\inst{\ref{aff59}}
\and C.~A.~J.~Duncan\orcid{0009-0003-3573-0791}\inst{\ref{aff9}}
\and X.~Dupac\inst{\ref{aff18}}
\and S.~Dusini\orcid{0000-0002-1128-0664}\inst{\ref{aff61}}
\and S.~Escoffier\orcid{0000-0002-2847-7498}\inst{\ref{aff62}}
\and M.~Farina\orcid{0000-0002-3089-7846}\inst{\ref{aff63}}
\and R.~Farinelli\inst{\ref{aff7}}
\and S.~Ferriol\inst{\ref{aff49}}
\and F.~Finelli\orcid{0000-0002-6694-3269}\inst{\ref{aff7},\ref{aff64}}
\and S.~Fotopoulou\orcid{0000-0002-9686-254X}\inst{\ref{aff65}}
\and M.~Frailis\orcid{0000-0002-7400-2135}\inst{\ref{aff21}}
\and E.~Franceschi\orcid{0000-0002-0585-6591}\inst{\ref{aff7}}
\and L.~Gabarra\orcid{0000-0002-8486-8856}\inst{\ref{aff66}}
\and S.~Galeotta\orcid{0000-0002-3748-5115}\inst{\ref{aff21}}
\and K.~George\orcid{0000-0002-1734-8455}\inst{\ref{aff67}}
\and B.~Gillis\orcid{0000-0002-4478-1270}\inst{\ref{aff9}}
\and C.~Giocoli\orcid{0000-0002-9590-7961}\inst{\ref{aff7},\ref{aff16}}
\and J.~Gracia-Carpio\inst{\ref{aff14}}
\and A.~Grazian\orcid{0000-0002-5688-0663}\inst{\ref{aff68}}
\and F.~Grupp\inst{\ref{aff14},\ref{aff69}}
\and W.~G.~Hartley\inst{\ref{aff59}}
\and S.~V.~H.~Haugan\orcid{0000-0001-9648-7260}\inst{\ref{aff70}}
\and W.~Holmes\inst{\ref{aff71}}
\and I.~M.~Hook\orcid{0000-0002-2960-978X}\inst{\ref{aff72}}
\and F.~Hormuth\inst{\ref{aff73}}
\and A.~Hornstrup\orcid{0000-0002-3363-0936}\inst{\ref{aff74},\ref{aff75}}
\and K.~Jahnke\orcid{0000-0003-3804-2137}\inst{\ref{aff76}}
\and M.~Jhabvala\inst{\ref{aff77}}
\and B.~Joachimi\orcid{0000-0001-7494-1303}\inst{\ref{aff78}}
\and E.~Keih\"anen\orcid{0000-0003-1804-7715}\inst{\ref{aff79}}
\and S.~Kermiche\orcid{0000-0002-0302-5735}\inst{\ref{aff62}}
\and A.~Kiessling\orcid{0000-0002-2590-1273}\inst{\ref{aff71}}
\and B.~Kubik\orcid{0009-0006-5823-4880}\inst{\ref{aff49}}
\and M.~K\"ummel\orcid{0000-0003-2791-2117}\inst{\ref{aff69}}
\and M.~Kunz\orcid{0000-0002-3052-7394}\inst{\ref{aff80}}
\and H.~Kurki-Suonio\orcid{0000-0002-4618-3063}\inst{\ref{aff81},\ref{aff82}}
\and A.~M.~C.~Le~Brun\orcid{0000-0002-0936-4594}\inst{\ref{aff83}}
\and S.~Ligori\orcid{0000-0003-4172-4606}\inst{\ref{aff33}}
\and P.~B.~Lilje\orcid{0000-0003-4324-7794}\inst{\ref{aff70}}
\and V.~Lindholm\orcid{0000-0003-2317-5471}\inst{\ref{aff81},\ref{aff82}}
\and I.~Lloro\orcid{0000-0001-5966-1434}\inst{\ref{aff84}}
\and M.~Magliocchetti\orcid{0000-0001-9158-4838}\inst{\ref{aff63}}
\and G.~Mainetti\orcid{0000-0003-2384-2377}\inst{\ref{aff85}}
\and D.~Maino\inst{\ref{aff86},\ref{aff36},\ref{aff87}}
\and E.~Maiorano\orcid{0000-0003-2593-4355}\inst{\ref{aff7}}
\and O.~Mansutti\orcid{0000-0001-5758-4658}\inst{\ref{aff21}}
\and S.~Marcin\inst{\ref{aff88}}
\and O.~Marggraf\orcid{0000-0001-7242-3852}\inst{\ref{aff89}}
\and M.~Martinelli\orcid{0000-0002-6943-7732}\inst{\ref{aff37},\ref{aff38}}
\and N.~Martinet\orcid{0000-0003-2786-7790}\inst{\ref{aff50}}
\and F.~Marulli\orcid{0000-0002-8850-0303}\inst{\ref{aff15},\ref{aff7},\ref{aff16}}
\and R.~J.~Massey\orcid{0000-0002-6085-3780}\inst{\ref{aff90}}
\and N.~Mauri\orcid{0000-0001-8196-1548}\inst{\ref{aff45},\ref{aff16}}
\and S.~Maurogordato\inst{\ref{aff91}}
\and C.~J.~R.~McPartland\orcid{0000-0003-0639-025X}\inst{\ref{aff75},\ref{aff92}}
\and E.~Medinaceli\orcid{0000-0002-4040-7783}\inst{\ref{aff7}}
\and S.~Mei\orcid{0000-0002-2849-559X}\inst{\ref{aff93},\ref{aff94}}
\and M.~Melchior\inst{\ref{aff95}}
\and Y.~Mellier\thanks{Deceased}\inst{\ref{aff11},\ref{aff96}}
\and M.~Meneghetti\orcid{0000-0003-1225-7084}\inst{\ref{aff7},\ref{aff16}}
\and E.~Merlin\orcid{0000-0001-6870-8900}\inst{\ref{aff37}}
\and G.~Meylan\inst{\ref{aff3}}
\and P.~Monaco\orcid{0000-0003-2083-7564}\inst{\ref{aff97},\ref{aff21},\ref{aff22},\ref{aff20}}
\and A.~Mora\orcid{0000-0002-1922-8529}\inst{\ref{aff98}}
\and M.~Moresco\orcid{0000-0002-7616-7136}\inst{\ref{aff15},\ref{aff7}}
\and C.~Moretti\orcid{0000-0003-3314-8936}\inst{\ref{aff21},\ref{aff20},\ref{aff22},\ref{aff23}}
\and R.~Nakajima\orcid{0009-0009-1213-7040}\inst{\ref{aff89}}
\and C.~Neissner\orcid{0000-0001-8524-4968}\inst{\ref{aff99},\ref{aff40}}
\and R.~C.~Nichol\orcid{0000-0003-0939-6518}\inst{\ref{aff100}}
\and S.-M.~Niemi\orcid{0009-0005-0247-0086}\inst{\ref{aff34}}
\and C.~Padilla\orcid{0000-0001-7951-0166}\inst{\ref{aff99}}
\and S.~Paltani\orcid{0000-0002-8108-9179}\inst{\ref{aff59}}
\and F.~Pasian\orcid{0000-0002-4869-3227}\inst{\ref{aff21}}
\and K.~Pedersen\inst{\ref{aff101}}
\and W.~J.~Percival\orcid{0000-0002-0644-5727}\inst{\ref{aff5},\ref{aff6},\ref{aff102}}
\and V.~Pettorino\orcid{0000-0002-4203-9320}\inst{\ref{aff34}}
\and A.~Pezzotta\orcid{0000-0003-0726-2268}\inst{\ref{aff19}}
\and S.~Pires\orcid{0000-0002-0249-2104}\inst{\ref{aff56}}
\and G.~Polenta\orcid{0000-0003-4067-9196}\inst{\ref{aff103}}
\and M.~Poncet\inst{\ref{aff104}}
\and L.~A.~Popa\inst{\ref{aff105}}
\and C.~Porciani\orcid{0000-0002-7797-2508}\inst{\ref{aff89}}
\and L.~Pozzetti\orcid{0000-0001-7085-0412}\inst{\ref{aff7}}
\and F.~Raison\orcid{0000-0002-7819-6918}\inst{\ref{aff14}}
\and R.~Rebolo\orcid{0000-0003-3767-7085}\inst{\ref{aff46},\ref{aff106},\ref{aff107}}
\and A.~Renzi\orcid{0000-0001-9856-1970}\inst{\ref{aff108},\ref{aff61}}
\and J.~Rhodes\orcid{0000-0002-4485-8549}\inst{\ref{aff71}}
\and G.~Riccio\inst{\ref{aff17}}
\and I.~Risso\orcid{0000-0003-2525-7761}\inst{\ref{aff19},\ref{aff26}}
\and E.~Romelli\orcid{0000-0003-3069-9222}\inst{\ref{aff21}}
\and M.~Roncarelli\orcid{0000-0001-9587-7822}\inst{\ref{aff7}}
\and R.~Saglia\orcid{0000-0003-0378-7032}\inst{\ref{aff69},\ref{aff14}}
\and Z.~Sakr\orcid{0000-0002-4823-3757}\inst{\ref{aff109},\ref{aff110},\ref{aff111}}
\and D.~Sapone\orcid{0000-0001-7089-4503}\inst{\ref{aff112}}
\and B.~Sartoris\orcid{0000-0003-1337-5269}\inst{\ref{aff69},\ref{aff21}}
\and J.~A.~Schewtschenko\orcid{0000-0002-4913-6393}\inst{\ref{aff9}}
\and M.~Schirmer\orcid{0000-0003-2568-9994}\inst{\ref{aff76}}
\and P.~Schneider\orcid{0000-0001-8561-2679}\inst{\ref{aff89}}
\and T.~Schrabback\orcid{0000-0002-6987-7834}\inst{\ref{aff113}}
\and A.~Secroun\orcid{0000-0003-0505-3710}\inst{\ref{aff62}}
\and G.~Seidel\orcid{0000-0003-2907-353X}\inst{\ref{aff76}}
\and S.~Serrano\orcid{0000-0002-0211-2861}\inst{\ref{aff114},\ref{aff115},\ref{aff2}}
\and P.~Simon\inst{\ref{aff89}}
\and C.~Sirignano\orcid{0000-0002-0995-7146}\inst{\ref{aff108},\ref{aff61}}
\and G.~Sirri\orcid{0000-0003-2626-2853}\inst{\ref{aff16}}
\and L.~Stanco\orcid{0000-0002-9706-5104}\inst{\ref{aff61}}
\and J.~Steinwagner\orcid{0000-0001-7443-1047}\inst{\ref{aff14}}
\and P.~Tallada-Cresp\'{i}\orcid{0000-0002-1336-8328}\inst{\ref{aff39},\ref{aff40}}
\and A.~N.~Taylor\inst{\ref{aff9}}
\and H.~I.~Teplitz\orcid{0000-0002-7064-5424}\inst{\ref{aff116}}
\and I.~Tereno\orcid{0000-0002-4537-6218}\inst{\ref{aff57},\ref{aff117}}
\and N.~Tessore\orcid{0000-0002-9696-7931}\inst{\ref{aff55}}
\and S.~Toft\orcid{0000-0003-3631-7176}\inst{\ref{aff118},\ref{aff92}}
\and R.~Toledo-Moreo\orcid{0000-0002-2997-4859}\inst{\ref{aff119}}
\and F.~Torradeflot\orcid{0000-0003-1160-1517}\inst{\ref{aff40},\ref{aff39}}
\and I.~Tutusaus\orcid{0000-0002-3199-0399}\inst{\ref{aff2},\ref{aff114},\ref{aff110}}
\and J.~Valiviita\orcid{0000-0001-6225-3693}\inst{\ref{aff81},\ref{aff82}}
\and T.~Vassallo\orcid{0000-0001-6512-6358}\inst{\ref{aff21}}
\and A.~Veropalumbo\orcid{0000-0003-2387-1194}\inst{\ref{aff19},\ref{aff26},\ref{aff25}}
\and Y.~Wang\orcid{0000-0002-4749-2984}\inst{\ref{aff116}}
\and J.~Weller\orcid{0000-0002-8282-2010}\inst{\ref{aff69},\ref{aff14}}
\and A.~Zacchei\orcid{0000-0003-0396-1192}\inst{\ref{aff21},\ref{aff20}}
\and G.~Zamorani\orcid{0000-0002-2318-301X}\inst{\ref{aff7}}
\and F.~M.~Zerbi\inst{\ref{aff19}}
\and I.~A.~Zinchenko\orcid{0000-0002-2944-2449}\inst{\ref{aff120}}
\and E.~Zucca\orcid{0000-0002-5845-8132}\inst{\ref{aff7}}
\and M.~Ballardini\orcid{0000-0003-4481-3559}\inst{\ref{aff121},\ref{aff122},\ref{aff7}}
\and E.~Bozzo\orcid{0000-0002-8201-1525}\inst{\ref{aff59}}
\and C.~Burigana\orcid{0000-0002-3005-5796}\inst{\ref{aff123},\ref{aff64}}
\and R.~Cabanac\orcid{0000-0001-6679-2600}\inst{\ref{aff110}}
\and M.~Calabrese\orcid{0000-0002-2637-2422}\inst{\ref{aff124},\ref{aff36}}
\and A.~Cappi\inst{\ref{aff91},\ref{aff7}}
\and T.~Castro\orcid{0000-0002-6292-3228}\inst{\ref{aff21},\ref{aff22},\ref{aff20},\ref{aff125}}
\and J.~A.~Escartin~Vigo\inst{\ref{aff14}}
\and J.~Garc\'ia-Bellido\orcid{0000-0002-9370-8360}\inst{\ref{aff126}}
\and R.~Maoli\orcid{0000-0002-6065-3025}\inst{\ref{aff127},\ref{aff37}}
\and J.~Mart\'{i}n-Fleitas\orcid{0000-0002-8594-569X}\inst{\ref{aff128}}
\and R.~B.~Metcalf\orcid{0000-0003-3167-2574}\inst{\ref{aff15},\ref{aff7}}
\and M.~P\"ontinen\orcid{0000-0001-5442-2530}\inst{\ref{aff81}}
\and V.~Scottez\orcid{0009-0008-3864-940X}\inst{\ref{aff11},\ref{aff129}}
\and M.~Sereno\orcid{0000-0003-0302-0325}\inst{\ref{aff7},\ref{aff16}}
\and M.~Tenti\orcid{0000-0002-4254-5901}\inst{\ref{aff16}}
\and M.~Viel\orcid{0000-0002-2642-5707}\inst{\ref{aff20},\ref{aff21},\ref{aff23},\ref{aff22},\ref{aff125}}
\and M.~Wiesmann\orcid{0009-0000-8199-5860}\inst{\ref{aff70}}
\and Y.~Akrami\orcid{0000-0002-2407-7956}\inst{\ref{aff126},\ref{aff130}}
\and I.~T.~Andika\orcid{0000-0001-6102-9526}\inst{\ref{aff69}}
\and G.~Angora\orcid{0000-0002-0316-6562}\inst{\ref{aff17},\ref{aff121}}
\and S.~Anselmi\orcid{0000-0002-3579-9583}\inst{\ref{aff61},\ref{aff108},\ref{aff131}}
\and M.~Archidiacono\orcid{0000-0003-4952-9012}\inst{\ref{aff86},\ref{aff87}}
\and F.~Atrio-Barandela\orcid{0000-0002-2130-2513}\inst{\ref{aff132}}
\and L.~Bazzanini\orcid{0000-0003-0727-0137}\inst{\ref{aff121},\ref{aff7}}
\and D.~Bertacca\orcid{0000-0002-2490-7139}\inst{\ref{aff108},\ref{aff68},\ref{aff61}}
\and M.~Bethermin\orcid{0000-0002-3915-2015}\inst{\ref{aff4}}
\and A.~Blanchard\orcid{0000-0001-8555-9003}\inst{\ref{aff110}}
\and L.~Blot\orcid{0000-0002-9622-7167}\inst{\ref{aff133},\ref{aff83}}
\and H.~B\"ohringer\orcid{0000-0001-8241-4204}\inst{\ref{aff14},\ref{aff67},\ref{aff134}}
\and M.~Bonici\orcid{0000-0002-8430-126X}\inst{\ref{aff5},\ref{aff36}}
\and S.~Borgani\orcid{0000-0001-6151-6439}\inst{\ref{aff97},\ref{aff20},\ref{aff21},\ref{aff22},\ref{aff125}}
\and M.~L.~Brown\orcid{0000-0002-0370-8077}\inst{\ref{aff47}}
\and S.~Bruton\orcid{0000-0002-6503-5218}\inst{\ref{aff135}}
\and A.~Calabro\orcid{0000-0003-2536-1614}\inst{\ref{aff37}}
\and B.~Camacho~Quevedo\orcid{0000-0002-8789-4232}\inst{\ref{aff20},\ref{aff23},\ref{aff21}}
\and F.~Caro\inst{\ref{aff37}}
\and C.~S.~Carvalho\inst{\ref{aff117}}
\and Y.~Charles\orcid{0009-0000-3636-834X}\inst{\ref{aff50}}
\and F.~Cogato\orcid{0000-0003-4632-6113}\inst{\ref{aff15},\ref{aff7}}
\and S.~Conseil\orcid{0000-0002-3657-4191}\inst{\ref{aff49}}
\and A.~R.~Cooray\orcid{0000-0002-3892-0190}\inst{\ref{aff136}}
\and M.~Costanzi\orcid{0000-0001-8158-1449}\inst{\ref{aff97},\ref{aff21},\ref{aff20}}
\and O.~Cucciati\orcid{0000-0002-9336-7551}\inst{\ref{aff7}}
\and S.~Davini\orcid{0000-0003-3269-1718}\inst{\ref{aff26}}
\and G.~Desprez\orcid{0000-0001-8325-1742}\inst{\ref{aff137}}
\and A.~D\'iaz-S\'anchez\orcid{0000-0003-0748-4768}\inst{\ref{aff138}}
\and S.~Di~Domizio\orcid{0000-0003-2863-5895}\inst{\ref{aff25},\ref{aff26}}
\and J.~M.~Diego\orcid{0000-0001-9065-3926}\inst{\ref{aff139}}
\and P.~Dimauro\orcid{0000-0001-7399-2854}\inst{\ref{aff140},\ref{aff37}}
\and P.-A.~Duc\orcid{0000-0003-3343-6284}\inst{\ref{aff4}}
\and M.~Y.~Elkhashab\orcid{0000-0001-9306-2603}\inst{\ref{aff21},\ref{aff22},\ref{aff97},\ref{aff20}}
\and A.~Enia\orcid{0000-0002-0200-2857}\inst{\ref{aff7}}
\and Y.~Fang\orcid{0000-0002-0334-6950}\inst{\ref{aff69}}
\and F.~Fontanot\orcid{0000-0003-4744-0188}\inst{\ref{aff21},\ref{aff20}}
\and A.~Franco\orcid{0000-0002-4761-366X}\inst{\ref{aff141},\ref{aff142},\ref{aff143}}
\and K.~Ganga\orcid{0000-0001-8159-8208}\inst{\ref{aff93}}
\and T.~Gasparetto\orcid{0000-0002-7913-4866}\inst{\ref{aff37}}
\and V.~Gautard\inst{\ref{aff144}}
\and R.~Gavazzi\orcid{0000-0002-5540-6935}\inst{\ref{aff50},\ref{aff96}}
\and E.~Gaztanaga\orcid{0000-0001-9632-0815}\inst{\ref{aff2},\ref{aff114},\ref{aff145}}
\and F.~Giacomini\orcid{0000-0002-3129-2814}\inst{\ref{aff16}}
\and F.~Gianotti\orcid{0000-0003-4666-119X}\inst{\ref{aff7}}
\and G.~Gozaliasl\orcid{0000-0002-0236-919X}\inst{\ref{aff146},\ref{aff81}}
\and M.~Guidi\orcid{0000-0001-9408-1101}\inst{\ref{aff24},\ref{aff7}}
\and C.~M.~Gutierrez\orcid{0000-0001-7854-783X}\inst{\ref{aff147}}
\and A.~Hall\orcid{0000-0002-3139-8651}\inst{\ref{aff9}}
\and H.~Hildebrandt\orcid{0000-0002-9814-3338}\inst{\ref{aff148}}
\and J.~Hjorth\orcid{0000-0002-4571-2306}\inst{\ref{aff101}}
\and L.~K.~Hunt\orcid{0000-0001-9162-2371}\inst{\ref{aff149}}
\and J.~J.~E.~Kajava\orcid{0000-0002-3010-8333}\inst{\ref{aff150},\ref{aff151},\ref{aff152}}
\and Y.~Kang\orcid{0009-0000-8588-7250}\inst{\ref{aff59}}
\and V.~Kansal\orcid{0000-0002-4008-6078}\inst{\ref{aff153},\ref{aff154}}
\and D.~Karagiannis\orcid{0000-0002-4927-0816}\inst{\ref{aff121},\ref{aff155}}
\and K.~Kiiveri\inst{\ref{aff79}}
\and J.~Kim\orcid{0000-0003-2776-2761}\inst{\ref{aff66}}
\and C.~C.~Kirkpatrick\inst{\ref{aff79}}
\and S.~Kruk\orcid{0000-0001-8010-8879}\inst{\ref{aff18}}
\and M.~Lattanzi\orcid{0000-0003-1059-2532}\inst{\ref{aff122}}
\and L.~Legrand\orcid{0000-0003-0610-5252}\inst{\ref{aff156},\ref{aff157}}
\and F.~Lepori\orcid{0009-0000-5061-7138}\inst{\ref{aff158}}
\and G.~Leroy\orcid{0009-0004-2523-4425}\inst{\ref{aff159},\ref{aff90}}
\and G.~F.~Lesci\orcid{0000-0002-4607-2830}\inst{\ref{aff15},\ref{aff7}}
\and J.~Lesgourgues\orcid{0000-0001-7627-353X}\inst{\ref{aff41}}
\and L.~Leuzzi\orcid{0009-0006-4479-7017}\inst{\ref{aff7}}
\and T.~I.~Liaudat\orcid{0000-0002-9104-314X}\inst{\ref{aff160}}
\and S.~J.~Liu\orcid{0000-0001-7680-2139}\inst{\ref{aff63}}
\and A.~Loureiro\orcid{0000-0002-4371-0876}\inst{\ref{aff161},\ref{aff162}}
\and J.~Macias-Perez\orcid{0000-0002-5385-2763}\inst{\ref{aff163}}
\and F.~Mannucci\orcid{0000-0002-4803-2381}\inst{\ref{aff149}}
\and C.~J.~A.~P.~Martins\orcid{0000-0002-4886-9261}\inst{\ref{aff164},\ref{aff28}}
\and L.~Maurin\orcid{0000-0002-8406-0857}\inst{\ref{aff60}}
\and M.~Miluzio\inst{\ref{aff18},\ref{aff165}}
\and A.~Montoro\orcid{0000-0003-4730-8590}\inst{\ref{aff2},\ref{aff114}}
\and G.~Morgante\inst{\ref{aff7}}
\and S.~Nadathur\orcid{0000-0001-9070-3102}\inst{\ref{aff145}}
\and K.~Naidoo\orcid{0000-0002-9182-1802}\inst{\ref{aff145}}
\and P.~Natoli\orcid{0000-0003-0126-9100}\inst{\ref{aff121},\ref{aff122}}
\and A.~Navarro-Alsina\orcid{0000-0002-3173-2592}\inst{\ref{aff89}}
\and S.~Nesseris\orcid{0000-0002-0567-0324}\inst{\ref{aff126}}
\and D.~Paoletti\orcid{0000-0003-4761-6147}\inst{\ref{aff7},\ref{aff64}}
\and F.~Passalacqua\orcid{0000-0002-8606-4093}\inst{\ref{aff108},\ref{aff61}}
\and K.~Paterson\orcid{0000-0001-8340-3486}\inst{\ref{aff76}}
\and A.~Pisani\orcid{0000-0002-6146-4437}\inst{\ref{aff62}}
\and D.~Potter\orcid{0000-0002-0757-5195}\inst{\ref{aff158}}
\and G.~W.~Pratt\inst{\ref{aff56}}
\and S.~Quai\orcid{0000-0002-0449-8163}\inst{\ref{aff15},\ref{aff7}}
\and M.~Radovich\orcid{0000-0002-3585-866X}\inst{\ref{aff68}}
\and G.~Rodighiero\orcid{0000-0002-9415-2296}\inst{\ref{aff108},\ref{aff68}}
\and W.~Roster\orcid{0000-0002-9149-6528}\inst{\ref{aff14}}
\and S.~Sacquegna\orcid{0000-0002-8433-6630}\inst{\ref{aff166}}
\and M.~Sahl\'en\orcid{0000-0003-0973-4804}\inst{\ref{aff167}}
\and D.~B.~Sanders\orcid{0000-0002-1233-9998}\inst{\ref{aff44}}
\and E.~Sarpa\orcid{0000-0002-1256-655X}\inst{\ref{aff23},\ref{aff125},\ref{aff22}}
\and J.~Schaye\orcid{0000-0002-0668-5560}\inst{\ref{aff35}}
\and A.~Schneider\orcid{0000-0001-7055-8104}\inst{\ref{aff158}}
\and M.~Schultheis\inst{\ref{aff91}}
\and D.~Sciotti\orcid{0009-0008-4519-2620}\inst{\ref{aff37},\ref{aff38}}
\and E.~Sellentin\inst{\ref{aff168},\ref{aff35}}
\and F.~Shankar\orcid{0000-0001-8973-5051}\inst{\ref{aff169}}
\and L.~C.~Smith\orcid{0000-0002-3259-2771}\inst{\ref{aff170}}
\and J.~G.~Sorce\orcid{0000-0002-2307-2432}\inst{\ref{aff171},\ref{aff60}}
\and S.~A.~Stanford\orcid{0000-0003-0122-0841}\inst{\ref{aff172}}
\and K.~Tanidis\orcid{0000-0001-9843-5130}\inst{\ref{aff66}}
\and F.~Tarsitano\orcid{0000-0002-5919-0238}\inst{\ref{aff173},\ref{aff59}}
\and G.~Testera\inst{\ref{aff26}}
\and R.~Teyssier\orcid{0000-0001-7689-0933}\inst{\ref{aff174}}
\and S.~Tosi\orcid{0000-0002-7275-9193}\inst{\ref{aff25},\ref{aff26},\ref{aff19}}
\and A.~Troja\orcid{0000-0003-0239-4595}\inst{\ref{aff108},\ref{aff61}}
\and M.~Tucci\inst{\ref{aff59}}
\and A.~Venhola\orcid{0000-0001-6071-4564}\inst{\ref{aff175}}
\and D.~Vergani\orcid{0000-0003-0898-2216}\inst{\ref{aff7}}
\and G.~Verza\orcid{0000-0002-1886-8348}$^{\star\star}$\inst{\ref{aff176},\ref{aff177}}
\and P.~Vielzeuf\orcid{0000-0003-2035-9339}\inst{\ref{aff62}}
\and S.~Vinciguerra\orcid{0009-0005-4018-3184}\inst{\ref{aff50}}
\and N.~A.~Walton\orcid{0000-0003-3983-8778}\inst{\ref{aff170}}}

\institute{School of Physics and Astronomy, University of Nottingham, University Park, Nottingham NG7 2RD, UK\label{aff1}
\and
Institute of Space Sciences (ICE, CSIC), Campus UAB, Carrer de Can Magrans, s/n, 08193 Barcelona, Spain\label{aff2}
\and
Institute of Physics, Laboratory of Astrophysics, Ecole Polytechnique F\'ed\'erale de Lausanne (EPFL), Observatoire de Sauverny, 1290 Versoix, Switzerland\label{aff3}
\and
Universit\'e de Strasbourg, CNRS, Observatoire astronomique de Strasbourg, UMR 7550, 67000 Strasbourg, France\label{aff4}
\and
Waterloo Centre for Astrophysics, University of Waterloo, Waterloo, Ontario N2L 3G1, Canada\label{aff5}
\and
Department of Physics and Astronomy, University of Waterloo, Waterloo, Ontario N2L 3G1, Canada\label{aff6}
\and
INAF-Osservatorio di Astrofisica e Scienza dello Spazio di Bologna, Via Piero Gobetti 93/3, 40129 Bologna, Italy\label{aff7}
\and
Departamento de F\'isica Te\'orica, Facultad de Ciencias, Universidad Aut\'onoma de Madrid, 28049 Cantoblanco, Madrid, Spain\label{aff8}
\and
Institute for Astronomy, University of Edinburgh, Royal Observatory, Blackford Hill, Edinburgh EH9 3HJ, UK\label{aff9}
\and
Centro de Investigaci\'{o}n Avanzada en F\'isica Fundamental (CIAFF), Facultad de Ciencias, Universidad Aut\'{o}noma de Madrid, 28049 Madrid, Spain\label{aff10}
\and
Institut d'Astrophysique de Paris, 98bis Boulevard Arago, 75014, Paris, France\label{aff11}
\and
OCA, P.H.C Boulevard de l'Observatoire CS 34229, 06304 Nice Cedex 4, France\label{aff12}
\and
Department of Astronomy, University of Florida, Bryant Space Science Center, Gainesville, FL 32611, USA\label{aff13}
\and
Max Planck Institute for Extraterrestrial Physics, Giessenbachstr. 1, 85748 Garching, Germany\label{aff14}
\and
Dipartimento di Fisica e Astronomia "Augusto Righi" - Alma Mater Studiorum Universit\`a di Bologna, via Piero Gobetti 93/2, 40129 Bologna, Italy\label{aff15}
\and
INFN-Sezione di Bologna, Viale Berti Pichat 6/2, 40127 Bologna, Italy\label{aff16}
\and
INAF-Osservatorio Astronomico di Capodimonte, Via Moiariello 16, 80131 Napoli, Italy\label{aff17}
\and
ESAC/ESA, Camino Bajo del Castillo, s/n., Urb. Villafranca del Castillo, 28692 Villanueva de la Ca\~nada, Madrid, Spain\label{aff18}
\and
INAF-Osservatorio Astronomico di Brera, Via Brera 28, 20122 Milano, Italy\label{aff19}
\and
IFPU, Institute for Fundamental Physics of the Universe, via Beirut 2, 34151 Trieste, Italy\label{aff20}
\and
INAF-Osservatorio Astronomico di Trieste, Via G. B. Tiepolo 11, 34143 Trieste, Italy\label{aff21}
\and
INFN, Sezione di Trieste, Via Valerio 2, 34127 Trieste TS, Italy\label{aff22}
\and
SISSA, International School for Advanced Studies, Via Bonomea 265, 34136 Trieste TS, Italy\label{aff23}
\and
Dipartimento di Fisica e Astronomia, Universit\`a di Bologna, Via Gobetti 93/2, 40129 Bologna, Italy\label{aff24}
\and
Dipartimento di Fisica, Universit\`a di Genova, Via Dodecaneso 33, 16146, Genova, Italy\label{aff25}
\and
INFN-Sezione di Genova, Via Dodecaneso 33, 16146, Genova, Italy\label{aff26}
\and
Department of Physics "E. Pancini", University Federico II, Via Cinthia 6, 80126, Napoli, Italy\label{aff27}
\and
Instituto de Astrof\'isica e Ci\^encias do Espa\c{c}o, Universidade do Porto, CAUP, Rua das Estrelas, PT4150-762 Porto, Portugal\label{aff28}
\and
Faculdade de Ci\^encias da Universidade do Porto, Rua do Campo de Alegre, 4150-007 Porto, Portugal\label{aff29}
\and
European Southern Observatory, Karl-Schwarzschild-Str.~2, 85748 Garching, Germany\label{aff30}
\and
Dipartimento di Fisica, Universit\`a degli Studi di Torino, Via P. Giuria 1, 10125 Torino, Italy\label{aff31}
\and
INFN-Sezione di Torino, Via P. Giuria 1, 10125 Torino, Italy\label{aff32}
\and
INAF-Osservatorio Astrofisico di Torino, Via Osservatorio 20, 10025 Pino Torinese (TO), Italy\label{aff33}
\and
European Space Agency/ESTEC, Keplerlaan 1, 2201 AZ Noordwijk, The Netherlands\label{aff34}
\and
Leiden Observatory, Leiden University, Einsteinweg 55, 2333 CC Leiden, The Netherlands\label{aff35}
\and
INAF-IASF Milano, Via Alfonso Corti 12, 20133 Milano, Italy\label{aff36}
\and
INAF-Osservatorio Astronomico di Roma, Via Frascati 33, 00078 Monteporzio Catone, Italy\label{aff37}
\and
INFN-Sezione di Roma, Piazzale Aldo Moro, 2 - c/o Dipartimento di Fisica, Edificio G. Marconi, 00185 Roma, Italy\label{aff38}
\and
Centro de Investigaciones Energ\'eticas, Medioambientales y Tecnol\'ogicas (CIEMAT), Avenida Complutense 40, 28040 Madrid, Spain\label{aff39}
\and
Port d'Informaci\'{o} Cient\'{i}fica, Campus UAB, C. Albareda s/n, 08193 Bellaterra (Barcelona), Spain\label{aff40}
\and
Institute for Theoretical Particle Physics and Cosmology (TTK), RWTH Aachen University, 52056 Aachen, Germany\label{aff41}
\and
Deutsches Zentrum f\"ur Luft- und Raumfahrt e. V. (DLR), Linder H\"ohe, 51147 K\"oln, Germany\label{aff42}
\and
INFN section of Naples, Via Cinthia 6, 80126, Napoli, Italy\label{aff43}
\and
Institute for Astronomy, University of Hawaii, 2680 Woodlawn Drive, Honolulu, HI 96822, USA\label{aff44}
\and
Dipartimento di Fisica e Astronomia "Augusto Righi" - Alma Mater Studiorum Universit\`a di Bologna, Viale Berti Pichat 6/2, 40127 Bologna, Italy\label{aff45}
\and
Instituto de Astrof\'{\i}sica de Canarias, E-38205 La Laguna, Tenerife, Spain\label{aff46}
\and
Jodrell Bank Centre for Astrophysics, Department of Physics and Astronomy, University of Manchester, Oxford Road, Manchester M13 9PL, UK\label{aff47}
\and
European Space Agency/ESRIN, Largo Galileo Galilei 1, 00044 Frascati, Roma, Italy\label{aff48}
\and
Universit\'e Claude Bernard Lyon 1, CNRS/IN2P3, IP2I Lyon, UMR 5822, Villeurbanne, F-69100, France\label{aff49}
\and
Aix-Marseille Universit\'e, CNRS, CNES, LAM, Marseille, France\label{aff50}
\and
Institut de Ci\`{e}ncies del Cosmos (ICCUB), Universitat de Barcelona (IEEC-UB), Mart\'{i} i Franqu\`{e}s 1, 08028 Barcelona, Spain\label{aff51}
\and
Instituci\'o Catalana de Recerca i Estudis Avan\c{c}ats (ICREA), Passeig de Llu\'{\i}s Companys 23, 08010 Barcelona, Spain\label{aff52}
\and
Institut de Ciencies de l'Espai (IEEC-CSIC), Campus UAB, Carrer de Can Magrans, s/n Cerdanyola del Vall\'es, 08193 Barcelona, Spain\label{aff53}
\and
UCB Lyon 1, CNRS/IN2P3, IUF, IP2I Lyon, 4 rue Enrico Fermi, 69622 Villeurbanne, France\label{aff54}
\and
Mullard Space Science Laboratory, University College London, Holmbury St Mary, Dorking, Surrey RH5 6NT, UK\label{aff55}
\and
Universit\'e Paris-Saclay, Universit\'e Paris Cit\'e, CEA, CNRS, AIM, 91191, Gif-sur-Yvette, France\label{aff56}
\and
Departamento de F\'isica, Faculdade de Ci\^encias, Universidade de Lisboa, Edif\'icio C8, Campo Grande, PT1749-016 Lisboa, Portugal\label{aff57}
\and
Instituto de Astrof\'isica e Ci\^encias do Espa\c{c}o, Faculdade de Ci\^encias, Universidade de Lisboa, Campo Grande, 1749-016 Lisboa, Portugal\label{aff58}
\and
Department of Astronomy, University of Geneva, ch. d'Ecogia 16, 1290 Versoix, Switzerland\label{aff59}
\and
Universit\'e Paris-Saclay, CNRS, Institut d'astrophysique spatiale, 91405, Orsay, France\label{aff60}
\and
INFN-Padova, Via Marzolo 8, 35131 Padova, Italy\label{aff61}
\and
Aix-Marseille Universit\'e, CNRS/IN2P3, CPPM, Marseille, France\label{aff62}
\and
INAF-Istituto di Astrofisica e Planetologia Spaziali, via del Fosso del Cavaliere, 100, 00100 Roma, Italy\label{aff63}
\and
INFN-Bologna, Via Irnerio 46, 40126 Bologna, Italy\label{aff64}
\and
School of Physics, HH Wills Physics Laboratory, University of Bristol, Tyndall Avenue, Bristol, BS8 1TL, UK\label{aff65}
\and
Department of Physics, University of Oxford, Keble Road, Oxford OX1 3RH, UK\label{aff66}
\and
University Observatory, LMU Faculty of Physics, Scheinerstr.~1, 81679 Munich, Germany\label{aff67}
\and
INAF-Osservatorio Astronomico di Padova, Via dell'Osservatorio 5, 35122 Padova, Italy\label{aff68}
\and
Universit\"ats-Sternwarte M\"unchen, Fakult\"at f\"ur Physik, Ludwig-Maximilians-Universit\"at M\"unchen, Scheinerstr.~1, 81679 M\"unchen, Germany\label{aff69}
\and
Institute of Theoretical Astrophysics, University of Oslo, P.O. Box 1029 Blindern, 0315 Oslo, Norway\label{aff70}
\and
Jet Propulsion Laboratory, California Institute of Technology, 4800 Oak Grove Drive, Pasadena, CA, 91109, USA\label{aff71}
\and
Department of Physics, Lancaster University, Lancaster, LA1 4YB, UK\label{aff72}
\and
Felix Hormuth Engineering, Goethestr. 17, 69181 Leimen, Germany\label{aff73}
\and
Technical University of Denmark, Elektrovej 327, 2800 Kgs. Lyngby, Denmark\label{aff74}
\and
Cosmic Dawn Center (DAWN), Denmark\label{aff75}
\and
Max-Planck-Institut f\"ur Astronomie, K\"onigstuhl 17, 69117 Heidelberg, Germany\label{aff76}
\and
NASA Goddard Space Flight Center, Greenbelt, MD 20771, USA\label{aff77}
\and
Department of Physics and Astronomy, University College London, Gower Street, London WC1E 6BT, UK\label{aff78}
\and
Department of Physics and Helsinki Institute of Physics, Gustaf H\"allstr\"omin katu 2, University of Helsinki, 00014 Helsinki, Finland\label{aff79}
\and
Universit\'e de Gen\`eve, D\'epartement de Physique Th\'eorique and Centre for Astroparticle Physics, 24 quai Ernest-Ansermet, CH-1211 Gen\`eve 4, Switzerland\label{aff80}
\and
Department of Physics, P.O. Box 64, University of Helsinki, 00014 Helsinki, Finland\label{aff81}
\and
Helsinki Institute of Physics, Gustaf H{\"a}llstr{\"o}min katu 2, University of Helsinki, 00014 Helsinki, Finland\label{aff82}
\and
Laboratoire d'etude de l'Univers et des phenomenes eXtremes, Observatoire de Paris, Universit\'e PSL, Sorbonne Universit\'e, CNRS, 92190 Meudon, France\label{aff83}
\and
SKAO, Jodrell Bank, Lower Withington, Macclesfield SK11 9FT, UK\label{aff84}
\and
Centre de Calcul de l'IN2P3/CNRS, 21 avenue Pierre de Coubertin 69627 Villeurbanne Cedex, France\label{aff85}
\and
Dipartimento di Fisica "Aldo Pontremoli", Universit\`a degli Studi di Milano, Via Celoria 16, 20133 Milano, Italy\label{aff86}
\and
INFN-Sezione di Milano, Via Celoria 16, 20133 Milano, Italy\label{aff87}
\and
University of Applied Sciences and Arts of Northwestern Switzerland, School of Computer Science, 5210 Windisch, Switzerland\label{aff88}
\and
Universit\"at Bonn, Argelander-Institut f\"ur Astronomie, Auf dem H\"ugel 71, 53121 Bonn, Germany\label{aff89}
\and
Department of Physics, Institute for Computational Cosmology, Durham University, South Road, Durham, DH1 3LE, UK\label{aff90}
\and
Universit\'e C\^{o}te d'Azur, Observatoire de la C\^{o}te d'Azur, CNRS, Laboratoire Lagrange, Bd de l'Observatoire, CS 34229, 06304 Nice cedex 4, France\label{aff91}
\and
Niels Bohr Institute, University of Copenhagen, Jagtvej 128, 2200 Copenhagen, Denmark\label{aff92}
\and
Universit\'e Paris Cit\'e, CNRS, Astroparticule et Cosmologie, 75013 Paris, France\label{aff93}
\and
CNRS-UCB International Research Laboratory, Centre Pierre Bin\'etruy, IRL2007, CPB-IN2P3, Berkeley, USA\label{aff94}
\and
University of Applied Sciences and Arts of Northwestern Switzerland, School of Engineering, 5210 Windisch, Switzerland\label{aff95}
\and
Institut d'Astrophysique de Paris, UMR 7095, CNRS, and Sorbonne Universit\'e, 98 bis boulevard Arago, 75014 Paris, France\label{aff96}
\and
Dipartimento di Fisica - Sezione di Astronomia, Universit\`a di Trieste, Via Tiepolo 11, 34131 Trieste, Italy\label{aff97}
\and
Telespazio UK S.L. for European Space Agency (ESA), Camino bajo del Castillo, s/n, Urbanizacion Villafranca del Castillo, Villanueva de la Ca\~nada, 28692 Madrid, Spain\label{aff98}
\and
Institut de F\'{i}sica d'Altes Energies (IFAE), The Barcelona Institute of Science and Technology, Campus UAB, 08193 Bellaterra (Barcelona), Spain\label{aff99}
\and
School of Mathematics and Physics, University of Surrey, Guildford, Surrey, GU2 7XH, UK\label{aff100}
\and
DARK, Niels Bohr Institute, University of Copenhagen, Jagtvej 155, 2200 Copenhagen, Denmark\label{aff101}
\and
Perimeter Institute for Theoretical Physics, Waterloo, Ontario N2L 2Y5, Canada\label{aff102}
\and
Space Science Data Center, Italian Space Agency, via del Politecnico snc, 00133 Roma, Italy\label{aff103}
\and
Centre National d'Etudes Spatiales -- Centre spatial de Toulouse, 18 avenue Edouard Belin, 31401 Toulouse Cedex 9, France\label{aff104}
\and
Institute of Space Science, Str. Atomistilor, nr. 409 M\u{a}gurele, Ilfov, 077125, Romania\label{aff105}
\and
Consejo Superior de Investigaciones Cientificas, Calle Serrano 117, 28006 Madrid, Spain\label{aff106}
\and
Universidad de La Laguna, Dpto. Astrof\'\i sica, E-38206 La Laguna, Tenerife, Spain\label{aff107}
\and
Dipartimento di Fisica e Astronomia "G. Galilei", Universit\`a di Padova, Via Marzolo 8, 35131 Padova, Italy\label{aff108}
\and
Institut f\"ur Theoretische Physik, University of Heidelberg, Philosophenweg 16, 69120 Heidelberg, Germany\label{aff109}
\and
Institut de Recherche en Astrophysique et Plan\'etologie (IRAP), Universit\'e de Toulouse, CNRS, UPS, CNES, 14 Av. Edouard Belin, 31400 Toulouse, France\label{aff110}
\and
Universit\'e St Joseph; Faculty of Sciences, Beirut, Lebanon\label{aff111}
\and
Departamento de F\'isica, FCFM, Universidad de Chile, Blanco Encalada 2008, Santiago, Chile\label{aff112}
\and
Universit\"at Innsbruck, Institut f\"ur Astro- und Teilchenphysik, Technikerstr. 25/8, 6020 Innsbruck, Austria\label{aff113}
\and
Institut d'Estudis Espacials de Catalunya (IEEC),  Edifici RDIT, Campus UPC, 08860 Castelldefels, Barcelona, Spain\label{aff114}
\and
Satlantis, University Science Park, Sede Bld 48940, Leioa-Bilbao, Spain\label{aff115}
\and
Infrared Processing and Analysis Center, California Institute of Technology, Pasadena, CA 91125, USA\label{aff116}
\and
Instituto de Astrof\'isica e Ci\^encias do Espa\c{c}o, Faculdade de Ci\^encias, Universidade de Lisboa, Tapada da Ajuda, 1349-018 Lisboa, Portugal\label{aff117}
\and
Cosmic Dawn Center (DAWN)\label{aff118}
\and
Universidad Polit\'ecnica de Cartagena, Departamento de Electr\'onica y Tecnolog\'ia de Computadoras,  Plaza del Hospital 1, 30202 Cartagena, Spain\label{aff119}
\and
Astronomisches Rechen-Institut, Zentrum f\"ur Astronomie der Universit\"at Heidelberg, M\"onchhofstr. 12-14, 69120 Heidelberg, Germany\label{aff120}
\and
Dipartimento di Fisica e Scienze della Terra, Universit\`a degli Studi di Ferrara, Via Giuseppe Saragat 1, 44122 Ferrara, Italy\label{aff121}
\and
Istituto Nazionale di Fisica Nucleare, Sezione di Ferrara, Via Giuseppe Saragat 1, 44122 Ferrara, Italy\label{aff122}
\and
INAF, Istituto di Radioastronomia, Via Piero Gobetti 101, 40129 Bologna, Italy\label{aff123}
\and
Astronomical Observatory of the Autonomous Region of the Aosta Valley (OAVdA), Loc. Lignan 39, I-11020, Nus (Aosta Valley), Italy\label{aff124}
\and
ICSC - Centro Nazionale di Ricerca in High Performance Computing, Big Data e Quantum Computing, Via Magnanelli 2, Bologna, Italy\label{aff125}
\and
Instituto de F\'isica Te\'orica UAM-CSIC, Campus de Cantoblanco, 28049 Madrid, Spain\label{aff126}
\and
Dipartimento di Fisica, Sapienza Universit\`a di Roma, Piazzale Aldo Moro 2, 00185 Roma, Italy\label{aff127}
\and
Aurora Technology for European Space Agency (ESA), Camino bajo del Castillo, s/n, Urbanizacion Villafranca del Castillo, Villanueva de la Ca\~nada, 28692 Madrid, Spain\label{aff128}
\and
ICL, Junia, Universit\'e Catholique de Lille, LITL, 59000 Lille, France\label{aff129}
\and
CERCA/ISO, Department of Physics, Case Western Reserve University, 10900 Euclid Avenue, Cleveland, OH 44106, USA\label{aff130}
\and
Laboratoire Univers et Th\'eorie, Observatoire de Paris, Universit\'e PSL, Universit\'e Paris Cit\'e, CNRS, 92190 Meudon, France\label{aff131}
\and
Departamento de F{\'\i}sica Fundamental. Universidad de Salamanca. Plaza de la Merced s/n. 37008 Salamanca, Spain\label{aff132}
\and
Center for Data-Driven Discovery, Kavli IPMU (WPI), UTIAS, The University of Tokyo, Kashiwa, Chiba 277-8583, Japan\label{aff133}
\and
Max-Planck-Institut f\"ur Physik, Boltzmannstr. 8, 85748 Garching, Germany\label{aff134}
\and
California Institute of Technology, 1200 E California Blvd, Pasadena, CA 91125, USA\label{aff135}
\and
Department of Physics \& Astronomy, University of California Irvine, Irvine CA 92697, USA\label{aff136}
\and
Kapteyn Astronomical Institute, University of Groningen, PO Box 800, 9700 AV Groningen, The Netherlands\label{aff137}
\and
Departamento F\'isica Aplicada, Universidad Polit\'ecnica de Cartagena, Campus Muralla del Mar, 30202 Cartagena, Murcia, Spain\label{aff138}
\and
Instituto de F\'isica de Cantabria, Edificio Juan Jord\'a, Avenida de los Castros, 39005 Santander, Spain\label{aff139}
\and
Observatorio Nacional, Rua General Jose Cristino, 77-Bairro Imperial de Sao Cristovao, Rio de Janeiro, 20921-400, Brazil\label{aff140}
\and
INFN, Sezione di Lecce, Via per Arnesano, CP-193, 73100, Lecce, Italy\label{aff141}
\and
Department of Mathematics and Physics E. De Giorgi, University of Salento, Via per Arnesano, CP-I93, 73100, Lecce, Italy\label{aff142}
\and
INAF-Sezione di Lecce, c/o Dipartimento Matematica e Fisica, Via per Arnesano, 73100, Lecce, Italy\label{aff143}
\and
CEA Saclay, DFR/IRFU, Service d'Astrophysique, Bat. 709, 91191 Gif-sur-Yvette, France\label{aff144}
\and
Institute of Cosmology and Gravitation, University of Portsmouth, Portsmouth PO1 3FX, UK\label{aff145}
\and
Department of Computer Science, Aalto University, PO Box 15400, Espoo, FI-00 076, Finland\label{aff146}
\and
 Instituto de Astrof\'{\i}sica de Canarias, E-38205 La Laguna; Universidad de La Laguna, Dpto. Astrof\'\i sica, E-38206 La Laguna, Tenerife, Spain\label{aff147}
\and
Ruhr University Bochum, Faculty of Physics and Astronomy, Astronomical Institute (AIRUB), German Centre for Cosmological Lensing (GCCL), 44780 Bochum, Germany\label{aff148}
\and
INAF-Osservatorio Astrofisico di Arcetri, Largo E. Fermi 5, 50125, Firenze, Italy\label{aff149}
\and
Department of Physics and Astronomy, Vesilinnantie 5, University of Turku, 20014 Turku, Finland\label{aff150}
\and
Finnish Centre for Astronomy with ESO (FINCA), Quantum, Vesilinnantie 5, University of Turku, 20014 Turku, Finland\label{aff151}
\and
Serco for European Space Agency (ESA), Camino bajo del Castillo, s/n, Urbanizacion Villafranca del Castillo, Villanueva de la Ca\~nada, 28692 Madrid, Spain\label{aff152}
\and
ARC Centre of Excellence for Dark Matter Particle Physics, Melbourne, Australia\label{aff153}
\and
Centre for Astrophysics \& Supercomputing, Swinburne University of Technology,  Hawthorn, Victoria 3122, Australia\label{aff154}
\and
Department of Physics and Astronomy, University of the Western Cape, Bellville, Cape Town, 7535, South Africa\label{aff155}
\and
DAMTP, Centre for Mathematical Sciences, Wilberforce Road, Cambridge CB3 0WA, UK\label{aff156}
\and
Kavli Institute for Cosmology Cambridge, Madingley Road, Cambridge, CB3 0HA, UK\label{aff157}
\and
Department of Astrophysics, University of Zurich, Winterthurerstrasse 190, 8057 Zurich, Switzerland\label{aff158}
\and
Department of Physics, Centre for Extragalactic Astronomy, Durham University, South Road, Durham, DH1 3LE, UK\label{aff159}
\and
IRFU, CEA, Universit\'e Paris-Saclay 91191 Gif-sur-Yvette Cedex, France\label{aff160}
\and
Oskar Klein Centre for Cosmoparticle Physics, Department of Physics, Stockholm University, Stockholm, SE-106 91, Sweden\label{aff161}
\and
Astrophysics Group, Blackett Laboratory, Imperial College London, London SW7 2AZ, UK\label{aff162}
\and
Univ. Grenoble Alpes, CNRS, Grenoble INP, LPSC-IN2P3, 53, Avenue des Martyrs, 38000, Grenoble, France\label{aff163}
\and
Centro de Astrof\'{\i}sica da Universidade do Porto, Rua das Estrelas, 4150-762 Porto, Portugal\label{aff164}
\and
HE Space for European Space Agency (ESA), Camino bajo del Castillo, s/n, Urbanizacion Villafranca del Castillo, Villanueva de la Ca\~nada, 28692 Madrid, Spain\label{aff165}
\and
INAF - Osservatorio Astronomico d'Abruzzo, Via Maggini, 64100, Teramo, Italy\label{aff166}
\and
Theoretical astrophysics, Department of Physics and Astronomy, Uppsala University, Box 516, 751 37 Uppsala, Sweden\label{aff167}
\and
Mathematical Institute, University of Leiden, Einsteinweg 55, 2333 CA Leiden, The Netherlands\label{aff168}
\and
School of Physics \& Astronomy, University of Southampton, Highfield Campus, Southampton SO17 1BJ, UK\label{aff169}
\and
Institute of Astronomy, University of Cambridge, Madingley Road, Cambridge CB3 0HA, UK\label{aff170}
\and
Univ. Lille, CNRS, Centrale Lille, UMR 9189 CRIStAL, 59000 Lille, France\label{aff171}
\and
Department of Physics and Astronomy, University of California, Davis, CA 95616, USA\label{aff172}
\and
Institute for Particle Physics and Astrophysics, Dept. of Physics, ETH Zurich, Wolfgang-Pauli-Strasse 27, 8093 Zurich, Switzerland\label{aff173}
\and
Department of Astrophysical Sciences, Peyton Hall, Princeton University, Princeton, NJ 08544, USA\label{aff174}
\and
Space physics and astronomy research unit, University of Oulu, Pentti Kaiteran katu 1, FI-90014 Oulu, Finland\label{aff175}
\and
Center for Computational Astrophysics, Flatiron Institute, 162 5th Avenue, 10010, New York, NY, USA\label{aff176}
\and
International Centre for Theoretical Physics (ICTP), Strada Costiera 11, 34151 Trieste, Italy\label{aff177}}    

\date{Received 31 March 2026 / Accepted 27 July 2026}

\abstract{Colour provides key information to constrain the stellar population of the intracluster light (ICL), the diffuse light within galaxy clusters. We derive radial colour profiles ($\YE-\HE$, $\JE-\HE$, $\IE-\HE$) of the brightest central galaxy (BCG) and ICL for 168 galaxy clusters from the Dark Energy Survey redMaPPer catalogue within the redshift range $0.1 < z < 0.75$ and halo mass $M_{\rm 200m} > 10^{14.05}\,M_\odot/h$, within the Euclid Quick Data Release (Q1). By stacking the light profiles measured using elliptical annuli on the BCG and ICL images, we reliably measure the ICL colour out to approximately $300\,\mathrm{kpc}$, over which we find a colour gradient, with the ICL becoming bluer than the BCG out to approximately $150\,\mathrm{kpc}$. Using stellar population synthesis (SPS) models, we find that near-infrared (NIR) colours are sensitive to metallicity, but not age. Thus, we measure a metallicity profile in the BCG and ICL that extends out to 300\,kpc for a statistical sample for the first time. While the colours, and the detection of a colour and metallicity gradient, are robust, we note that our metallicity measurement is dependent upon our choice of both SPS model and colour. We find that the metallicity of the BCG is super-solar [$\logten (Z/Z_\odot) > 0.1$], while the ICL is metal-poor [$\logten(Z/Z_\odot)\approx -0.5$], placing the first statistical constraint on the metallicity of the ICL out to large radii. These results support our current understanding of BCG and ICL formation: BCGs form predominantly through mergers with massive galaxies or the cores of satellite galaxies, while the ICL grows via the tidal stripping of lower-metallicity stars from the outskirts of massive satellite galaxies.}


    \titlerunning{ICL metallicity}
    \authorrunning{Euclid Collaboration: J. B. Golden-Marx et al.}
    \maketitle
    
\section{Introduction}\label{sec:intro}

Galaxy clusters are massive galactic megacities that contain hundreds to thousands of galaxies located within an underlying structure of dark matter, referred to as the dark matter halo. Near the centre of each galaxy cluster is the brightest central galaxy (BCG), which is generally a massive, radially extended, elliptical galaxy, typically with little star formation, except for the case of cooling flow clusters \citep[e.g.][]{schombert86, lin04, vonderlinden07, lauer14}. 

Unlike similarly massive galaxies, BCGs are particularly radially extended \citep[e.g.][]{hoessel80, lynam00} and surrounded by a faint diffuse halo of light commonly referred to as intracluster light \citep[ICL;][]{zwicky33, zwicky51}. Both stacked observations \citep{zibetti05, zhang18, chen22, ahad25} and targeted deeper individual cluster observations with long exposure times \citep{dressler79, gonzalez21} have found that the ICL extends to megaparsec scales. As ICL observations become more common, we have been able to study the ICL as a statistical population \citep[e.g.][]{kluge20, golden-marx23, Golden-Marx25} and constrain its assembly.

Understanding how the ICL assembles is crucial because it encodes both the recent and long-term growth of galaxy clusters in ways not captured by galaxy light alone or using observables linked to the intracluster medium. The ICL preserves spatial, kinematic, and chemical signatures that record when mass was accreted, and from which progenitors, and along which directions, while also providing information about the overall gravitational potential of the cluster \citep{montes19}. Among these signatures, the metallicity and age of the ICL are especially informative, constraining both formation channels and progenitor populations. 

Metallicity encodes the chemical enrichment history of a stellar population and is linked to the stellar mass and star-formation history of the galaxies from which the stars were formed \citep{Gallazzi2005,Kirby2013}. Since galaxies have radial metallicity gradients \citep[e.g.][]{Davies93, kobayashi99, baes07, Lian2018, zhuang2019}, stars stripped from the outskirts of massive, metal-rich satellite galaxies will imprint a low metallicity on the ICL, as will stars from low-mass, metal-poor dwarf galaxies. In contrast, if the metal-rich cores of massive satellite galaxies merge with the ICL, they will imprint a high metallicity on the ICL \citep[e.g.][]{murante07,contini14, contini19}. Examining how the metallicity of the BCG and ICL varies with radius therefore clarifies the progenitor populations and provides a key observable for testing competing models of ICL formation, which we address in this analysis. Models dominated by the tidal stripping of satellite galaxies result in a metallicity gradient \citep{contini19, montenegro-taborda2023, brown2024}, whereas those driven by mergers with the BCG mix the stellar population and flatten the metallicity gradient \citep{contini19, montes22}.

In the last two decades, cosmological hydrodynamical simulations (e.g. IllustrisTNG, EAGLE) and semi-analytic models have predicted that the ICL should be systematically more metal-poor than the BCG at all epochs and detect a negative metallicity gradient, such that when measured radially outwards from the BCG, the metallicity decreases \citep[e.g.][]{contini14, contini19, Cui14, ahvazi24, Mayes26}. Moreover, these simulations and models predict ICL assembling earlier from low-mass satellites and later from more intermediate-mass systems as clusters grow \citep{murante04, rudick11, contini14}. Specifically, recent semi-analytic work finds that the typical progenitors of the ICL shift from $\logten (M_\star/M_\odot) \approx 9\text{--}10$ at early times to $\logten (M_\star/M_\odot) \approx 10\text{--}11$ at $z \lesssim 1$, naturally producing negative colour and metallicity gradients in the combined BCG and ICL (BCG+ICL) systems due to dynamical friction, such that the more massive galaxies fall farther into the core of the cluster before being stripped \citep{contini19}. As a result of these simulation studies, there are several predictions about the metallicity of the ICL that can be addressed using observational data: (i) the ICL (at ${\approx}\,100$\,kpc) should be ${\approx}\,0.2\text{--}0.4$\,dex more metal-poor than the BCG \citep{contini19, ahad23, Mayes26}; (ii) the BCG+ICL metallicity gradient should be negative and persist to $\gtrsim 200$\,kpc \citep{ahad23, ahvazi24, Mayes26}; (iii) the characteristic ICL metallicity should evolve mildly with redshift as the dominant progenitor mass scale increases after $z \approx 1$ \citep{murante07,contini14}; and (iv) clusters with a larger contribution from stripped low-mass satellites should show steeper (more negative) metallicity gradients than systems where the ICL is built predominantly via major mergers with the BCG, a trend testable via joint gradients and satellite galaxy mass functions \citep{contini14,contini19}. Confronting these predictions with spatially resolved metallicities from deep imaging provides a stringent check on ICL assembly pathways and their redshift evolution. Since ICL can account for between 10\% and 35\% of the stellar mass within a cluster \citep[e.g.][]{rudick11, Joo2023, ragusa23, jimenez-teja24, Golden-Marx25}, incorporating its metallicity is also essential for estimating the chemical-enrichment budget of the intracluster medium. 

The simulation predictions for the metallicity of the ICL have become increasingly testable over the last decade, in part because photometric colours can be used to infer the metallicity of the ICL. Both stacked measurements \citep{zibetti05, zhang18, zhang24} and individual observations \citep[e.g.][]{Montes2014, montes18, montes21, demaio15, demaio18, morishita17, ragusa21, golden-marx23} have found that the core of the BCG is redder than the outskirts, which has led some prior observational studies to suggest that the ICL (or more simply the BCG outskirts) is likely metal-poor compared to the BCG core \citep{loubser12, mcdermid2015, oliva-altamirano15, Johnston2018, ragusa2022}. However, in optical photometry a degeneracy exists between the metallicity and the age of the stellar population, which makes it unclear whether this colour gradient results from a metallicity gradient, an age gradient, or a combination of the two in the stellar population of the ICL. \textit{Hubble} Space Telescope (HST) studies of the Frontier Fields clusters find average ICL metallicities of $\logten (Z/Z_\odot) \approx -0.3$ to $-0.6$ and clear negative gradients from super-solar values in the BCG-dominated regions to sub-solar values at larger radii \citep{Montes2014, morishita17, demaio18, montes18}, while spectroscopic work in nearby clusters (e.g. Coma and $z \approx 0.05$ systems) report predominantly older, metal-poor ICL consistent with stripping from lower-mass progenitors \citep{melnick2012,edwards2016,edwards20,edwards24, Johnston2018, gu20}. More recently, deep optical/near-infrared (NIR) observations extend these trends to hundreds of kiloparsecs, with colour-to-metallicity conversions indicating gradients from roughly solar inside ${\approx}\,10$\,kpc to $\logten (Z/Z_\odot) \approx -0.7$ by ${\approx}\,200\text{--}300$\,kpc in the ICL-dominated regime \citep{ellien25, EROPerseusICL}. 

Beyond tracing ICL assembly, a well-constrained ICL metallicity, along with the estimated age of the stellar population, can reduce the uncertainty in the stellar mass estimates of the BCG and ICL. The stellar mass of the ICL strongly correlates with the cluster halo mass \citep{zibetti05, huang21, golden-marx23, Golden-Marx25, contreras-santos24} and simulations have shown that the radial density profile of dark matter in the halo scales with that of the ICL \citep{alonso_asensio2020, contreras-santos24, yoo24, butler2025}. Thus, constraining ICL metallicity and the age of the stellar population will also optimise the stellar mass measurement and the utility of the ICL as a tracer of the dark matter halo.  

Here, we present an analysis using the Euclid Quick Data Release \citep[Q1;][]{Q1cite} observations of ICL measurements in 168 galaxy clusters with $\logten (M_{\rm halo}/M_\odot/h) > 14.05$ over the redshift range $0.1 < z < 0.75$ identified using the redMaPPer cluster finding algorithm as part of the Dark Energy Survey \citep[DES;][]{rykoff14, DESY3}. We chose the DES-redMaPPer cluster catalogue because it provides BCG identification, which is vital to our ICL measurements. However, we note that our results are not dependent on our choice of catalogue and future analyses could be done using \Euclid cluster catalogues, which will extend out to much higher redshifts \citep{Q1-SP050}. In this analysis, we use the combined NIR and optical data from \Euclid to measure the colours of the ICL and examine how these colours vary as a function of redshift and radial extent. The remaining sections are as follows. In Sect.~\ref{sec:Data}, we discuss the Q1 data and the DES-redMaPPer cluster catalogue used in this analysis. In Sect.~\ref{sec:methods}, we discuss how we measure the flux and colour of the ICL. In Sect.~\ref{sec:results}, we present the colours of the ICL. In Sect.~\ref{sec:NIR}, we discuss what the NIR colours can tell us about the formation redshift and metallicity of a stellar population. In Sect.~\ref{sec:Discussion}, we discuss how the colours of the ICL relate to the metallicity and age of the stellar population. We conclude in Sect.~\ref{sec:conclusion}. 

In this analysis, we assume a flat $\Lambda$ cold dark matter Universe, with $\Omega_{\rm m}=0.30$, $\Omega_\Lambda=0.70$, and $H_0 = 100\,h\,\mathrm{km}\,\mathrm{s}^{-1}\,\mathrm{Mpc}^{-1}$, with $h = 0.7$. Additionally, throughout this analysis we assume a solar metallicity of 0.019 \citep{vazdekis2016} and note that all of the radial extent measurements for the annuli of the ICL are in physical distances. 

\section{Data} \label{sec:Data}
\subsection{\Euclid Q1 data}\label{subsec:Euclid_data}
\Euclid is a 1.2\,m telescope with a 0.54\,deg$^2$ field of view \citep{EuclidSkyOverview}. The faint detection limit, design of the wide-field telescope, and tight control on scattered light make \Euclid ideal for studying the low-surface-brightness (LSB) Universe \citep{Scaramella-EP1,Borlaff-EP16, EROData, EROPerseusDGs, Mondelin25, Saifollahi25, Urbano24}, in particular the ICL \citep{bellhouse25, ellien25, EROPerseusICL}. The Q1 data release \citep{Q1-TP001} consists of ${\approx}\,63.1$\,deg$^2$ in three deep fields: Euclid Deep Field North (EDF-N), Euclid Deep Field South (EDF-S), and Euclid Deep Field Fornax (EDF-F). Each field was observed in a single observation taken to the depth of the Euclid Wide Survey \citep{Scaramella-EP1}. The Q1 observations were taken in four photometric wavebands (\IE{}, \YE{}, \JE{}, and \HE{}) using the VIS \citep{EuclidSkyVIS} and NISP \citep{EuclidSkyNISP} instruments. The data used in this analysis were made publicly available as part of the Q1 data release \citep{Q1-TP001}.\footnote{The Q1 data are available at: \url{https://www.cosmos.esa.int/en/web/euclid/euclid-q1-data-release}} 

In this analysis, we focus only on the data from EDF-S and EDF-F, which are 28.1\,deg$^2$ and 12.1\,deg$^2$, respectively, and are covered entirely by DES. To obtain images that allow us to recover ICL signal, we followed the process described in \citet{Q1-Bamford}. We began with the Level 2 (LE2) single-exposure products generated by the Science Ground Segment (SGS) VIS and NIR pipelines \citep{Q1-TP002,Q1-TP003}. The calibrated exposures were then post-processed including the removal of the residual atemporal background signal, correction of time-dependent instrumental artefacts (e.g. `tartan' patterns), masking of persistence features (only for NIR images), resampling of individual exposures to a common pixel scale ($0.3\,\mathrm{arcsec\,pixel^{-1}}$), background matching across detector chips, and coaddition of the resampled exposures into a $30\arcmin \times 30\arcmin$ image for each cluster in the \IE, \YE, \JE, and \HE bands. In addition, we produce a coadded $YJ$\HE image by combining the \YE, \JE, and \HE images weighted by their relative bandwidths. This additional processing aims to suppress large-scale instrumental background structures while preserving faint extended emission. We note that reprocessing data for LSB observations is a common practice when using large cosmological surveys, such as SDSS and DES. These surveys have different scientific goals than LSB science, so they generally apply a local background subtraction that can remove the diffuse light surrounding bright galaxies \citep[e.g.][]{bernardi07, vonderlinden07, zhang19}. 

\subsection{DES-redMaPPer cluster sample}
In this analysis, we use the DES Y3 redMaPPer cluster catalogue version 6.4.22+2 \citep{DESY3},\footnote{The publicly available redMaPPer catalogues are found at \url{https://des.ncsa.illinois.edu/releases/y3a2/Y3key-cluster}} based on DES Year 3 Gold photometric data \citep{sevilla-noarbe21}. This version of the redMaPPer cluster catalogue contains over 21\,000 galaxy clusters with a redMaPPer membership probability-weighted total richness, $\lambda$, greater than 20 \citep{rykoff12, rykoff14}, which corresponds to a halo mass lower limit of ${\approx}\,10^{14.05}\,M_\odot/h$ \citep{DESY3}. 

redMaPPer is a probabilistic cluster finding algorithm that identifies galaxy clusters and their member galaxies as overdensities in the colour--magnitude phase space. The photometric redshift provided by redMaPPer for each cluster is determined based on a fit to the red sequence. redMaPPer aims to select the luminous cluster galaxy located nearest to the centre of the gravitational potential well of the cluster as the central galaxy and provides the information for the five most probable central candidates per cluster \citep{rykoff14, rykoff16}. We assume that the most likely central galaxy candidate is the BCG, although we visually confirm each of our BCGs. Based on multi-wavelength measurements, redMaPPer identifies the correct BCG in ${\approx}\,80\%$ of all clusters \citep{saro15, zhang19, bleem20, kelly24}. We note that \citet{Golden-Marx25} showed that BCG+ICL statistics are not biased by the inclusion of low-probability central galaxies.  

\begin{figure}
    \centering
    \includegraphics[width=9cm]{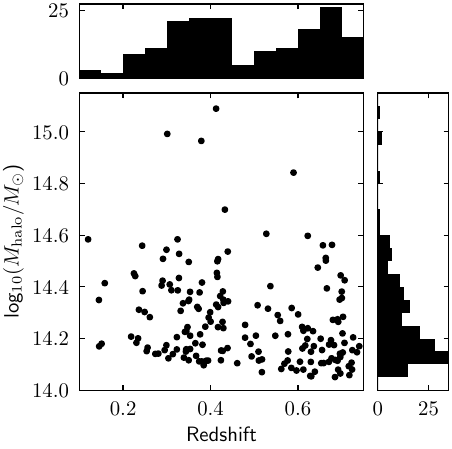}
    \caption{Halo mass, where $M_{\rm halo}$ refers to $M_{\rm 200m}$, of the redMaPPer clusters, estimated using the mass-richness relation from \citet{DESY3}, plotted against the redMaPPer photometric redshift for the 168 DES-redMaPPer clusters in our sample. The side panels are histograms of the halo mass and redshift, respectively.}
    \label{fig:RMinfo}
\end{figure}

Additionally, we note that \citet{zhang24, zhang26} and \citet{Golden-Marx25} use the same cluster catalogue for their analysis of the ICL. However, because we are using the Q1 photometry, here we are limited to the overlap region between the Q1 (EDF-S and EDF-F) and DES fields, which leaves us with a subset of 168 clusters covering the redshift range $0.1 < z < 0.75$. In Fig.~\ref{fig:RMinfo}, we plot the halo mass distribution, where $M_{\rm halo}$ refers to $M_{\rm 200m}$, estimated using the mass-richness relation given in appendix D from \citet{DESY3}, as a function of redshift. We note that the halo mass distribution of our sample is relatively uniform across the redshift range of interest. However, the redshift distribution of our sample has a deficit of clusters at $z \approx 0.5$, which results from limitations in the redMaPPer cluster catalogue. Upcoming cluster catalogues from Euclid Data Release 1 (DR1) data will not have these limitations and will therefore be free of the possible bias that the cluster sample selection might have on our results. 

\section{ICL measurement pipeline} \label{sec:methods}
\subsection{Background subtraction and masking}
The first step of our analysis was to measure and subtract the background flux within our BCG+ICL images. For this, we constructed a background mask, which identifies and masks all sources in the image, which includes bright and faint satellite galaxies, bright foreground stars, artefacts, and a conservative mask of the region where the BCG and ICL fluxes are significant. In this mask, high-surface-brightness sources located outside of the BCG+ICL region were grown around each detected source by 50\% to minimise the contamination of ICL from their extended halos. Within the BCG+ICL masked region, the masks are more conservatively grown by 25\%, to retain as much diffuse light as possible while excluding bright compact sources. In all NIR-band images, we used the source masks generated from the coadded $YJ$\HE image, while a separate source mask was used for the \IE-band image.

After applying this mask, we used \texttt{photutils.Background2D} to detect any large-scale gradients in our background. To measure the 2D background substructure within the image, we used a variable box size of 1\,Mpc $\times$ 1\,Mpc at the redshift of the cluster. The resulting background map was first median-filtered using a $3\times3$ grid to mitigate overestimation caused by any residual ICL contamination, which can extend beyond 1\,Mpc from the BCG, and then subtracted from our image to create our measurement image. 

We unmasked the region of BCG+ICL to construct masks that allow us to measure the BCG+ICL photometry. Since we combined NIR and optical photometry for the \Euclid colours, we combined the $YJ$\HE mask with the \IE{} mask to create a more conservative $IYJ$\HE measurement mask, which removes bright sources and potential artefacts from our BCG+ICL signal.

\subsection{Construction of the surface brightness profiles}
\label{subsec:SB}
The key measurement in this analysis is the colour of the ICL. For this colour to be physically meaningful, we required each photometric measurement to cover the same spatial region and have the same measurement mask applied (the final measurement mask). To construct the surface brightness (SB) profiles, we used the coadded $YJ$\HE image to extract the \texttt{AutoProf} isophotal shape profiles of the BCG+ICL \citep{Stone2021b}. The isophotes were fixed to have semi-major axes starting at 1\,pixel and increase in radius by 0.15\,dex for subsequent ellipses. 

To run \texttt{AutoProf}, we first provided a fixed centre, given by the DES-redMaPPer cluster catalogue. \texttt{AutoProf} applies the measurement mask ($IYJ$\HE mask) to the image and then estimates the position angle (PA) and ellipticity of each isophote using a Fourier analysis of the flux within circular apertures. Starting from the initial apertures, for each isophote, the PA and ellipticity are iteratively refined using a regularisation approach, which continues until convergence is achieved \citep{Stone2021b}. Thus, the final output is a series of isophote profiles with ellipticity and PA, computed out to where the signal-to-noise ratio (S/N) drops to 1. Beyond this radius the ellipticity and PA of the ellipses are fixed to the last robustly measured values. The typical accuracy of the PA and ellipticity is \ang{10} and 0.1, respectively \citep{Q1-Bamford}. To get our final photometric measurements, we used the elliptical isophotes defined by \texttt{AutoProf} to extract consistent 1D SB profiles across all four wavebands by measuring the median sigma-clipped flux within each annulus.

\subsection{ICL flux measurements and uncertainties}
\label{subsec:ICL_flux}
Using the SB profiles, we then integrated the 1D profiles (assuming an elliptical shape based on the \texttt{AutoProf} isophotal annuli) to measure the flux contained within each annulus. We then interpolated over those fluxes to estimate the total flux contained in elliptical annuli measured in fixed physical sizes as opposed to fixed angular sizes to compare clusters across redshift space. Based on the red sequence estimated photometric redshifts from redMaPPer, we measured the flux in the following physical annuli: 0--10\,kpc, 10--30\,kpc, 30--50\,kpc, 50--100\,kpc, 100--150\,kpc, 150--200\,kpc, 200--300\,kpc, 300--400\,kpc, 400--500\,kpc, and 500--600\,kpc. This process was then repeated for each cluster. We did not apply a Galactic extinction correction to our flux (or colour) measurements because, as shown in \citet{Q1-TP005}, the extinction corrections are less than 0.05 in magnitude for EDF-S and EDF-F, which is smaller than the uncertainty in our measurements.

To estimate the uncertainty of our flux measurement within each of our selected apertures, we performed the previously described analysis on two \ang{2.5}$\times$ \ang{2.5} sky patches, one for each deep field. We used the same background box size for the sky patch as for each cluster, so each cluster has a bespoke uncertainty in each annulus. After subtracting the background, for each field we measured the photometry centred on 1000 random points within each sky patch, with elliptical apertures (with an ellipticity of 0.5) extending out to 1000\arcsec\, with the same spacing in arcseconds as our ICL measurements as described in \citet{Q1-Bamford}. As for the ICL measurements, we interpolated these measurements to measure the light within the same physical annuli as described in the previous section. Using the 1000 SB profiles, we then measured the standard deviation within each radial bin to determine the background noise levels for each physical annulus. We note that the noise properties on large scales are dominated by systematic background variation, not Poisson noise, and therefore the additional masked area due to the satellite galaxies in each cluster, relative to the control field, does not impact our uncertainty estimate. 

\section{Colour of the BCG+ICL} \label{sec:results}

\begin{figure*}
    \centering
    \includegraphics[width=18cm,trim={0cm 0 0cm 0}]{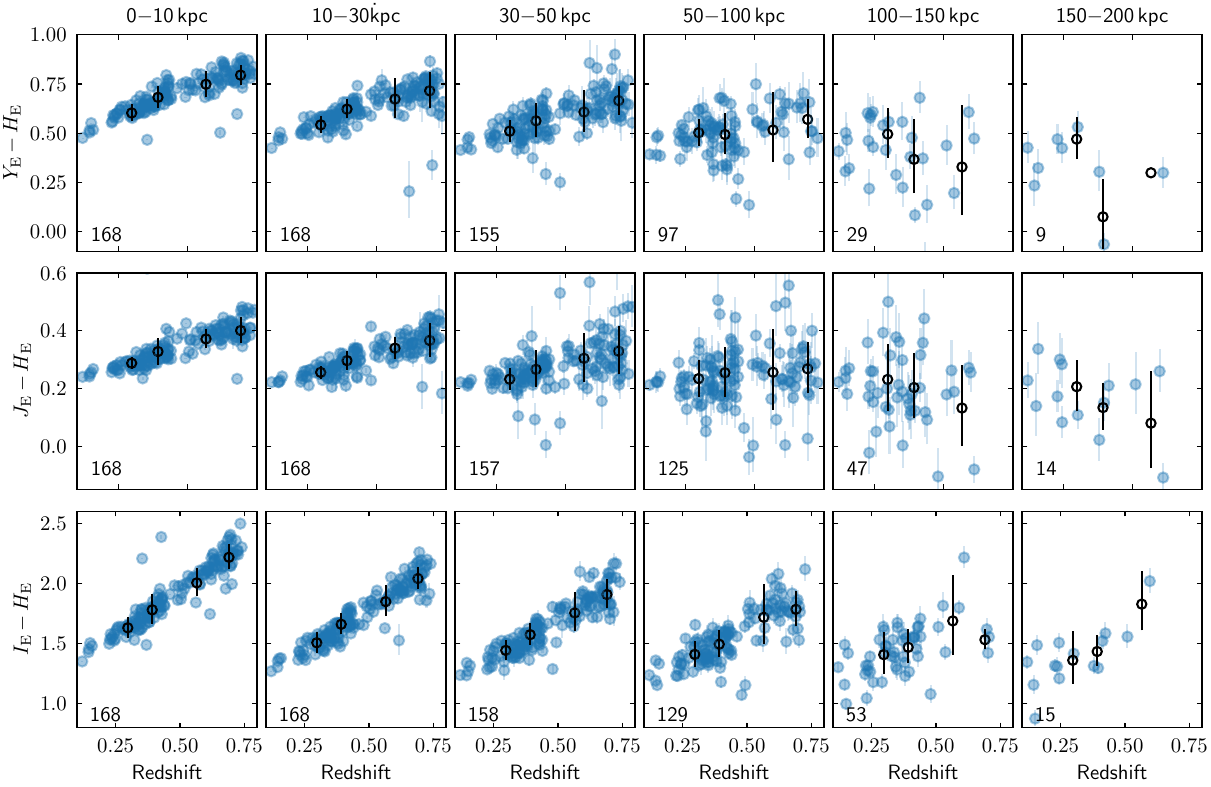}
    \caption{Colour measurements and their associated uncertainty for the first six apertures arranged in columns, with an increasing outer radius from left to right, and three different colours arranged in rows. We also overlay unfilled circles, which represent the weighted mean of all data above the S/N $>$ 10 limit (which includes ${<}\,10$\% of the sample at large radii). The number of clusters above the S/N $>$ 10 limit is given in the lower left corner of each subplot.}
    \label{fig:Colour_bin_plot}
\end{figure*}

\subsection{BCG+ICL colour in individual clusters}
As previously described, for each of the DES-redMaPPer clusters, we measured the flux as well as the associated uncertainty in the four wavebands and three colours ($\YE-\HE$, $\JE-\HE$, and $\IE-\HE$) in the ten non-overlapping physical apertures. Quantitatively, these colours are the ratio of the two fluxes, and the uncertainty in each colour was estimated using standard error propagation based on the measured flux ratio and the uncertainty in each respective flux. 

In Fig.~\ref{fig:Colour_bin_plot} we present the colour as a function of redshift for the first six radial apertures. To be included in this figure, we required that the S/N of the flux in both bands used to measure the colour be ${>}\,10$, which is why we excluded the outer four bins as there are very few clusters that meet this criterion. We provide the number of clusters that meet these criteria in the lower left corner of each sub-plot of Fig.~\ref{fig:Colour_bin_plot}. 

We can robustly measure the colour within 50\,kpc for ${>}\,90$\% of all the clusters. However, the number of clusters for which we can measure robust colours decreases significantly out to larger radii and is less than 10\% by 200\,kpc. We note that at such large radii the ICL is preferentially detectable among both higher-halo-mass and lower-redshift clusters, in agreement with some of the predictions from \citet{bellhouse25}. To enhance our understanding of whether there is any trend between the colours, we take a weighted mean of the photometric colour within four redshift bins (with median redshifts of 0.30, 0.39, 0.56, 0.69), shown as unfilled circles. The black error bars represent the 16th and 84th percentiles of the colour measurements within the redshift range at that aperture, and therefore highlight the scatter in the BCG+ICL colour between individual clusters at a certain redshift. We see a strong consistency in the colours of the BCGs within each redshift bin within the inner 30\,kpc. This implies that the stellar populations of the BCGs within these different clusters are likely to be similar. However, at larger extents, we see a larger scatter in the colours of the ICL in different clusters within the same redshift bin. The scatter in colours is significantly larger than the individual error bars on the colour for each cluster. This spread in colour implies that the stellar population of the ICL differs from cluster to cluster, which suggests that individual clusters, or small cluster samples, should not be used to characterise the average properties of the ICL population.

\subsection{Average BCG+ICL colour}

\begin{figure*}
    \centering
    \includegraphics[width=18cm,trim={0cm 0 0cm 0}]{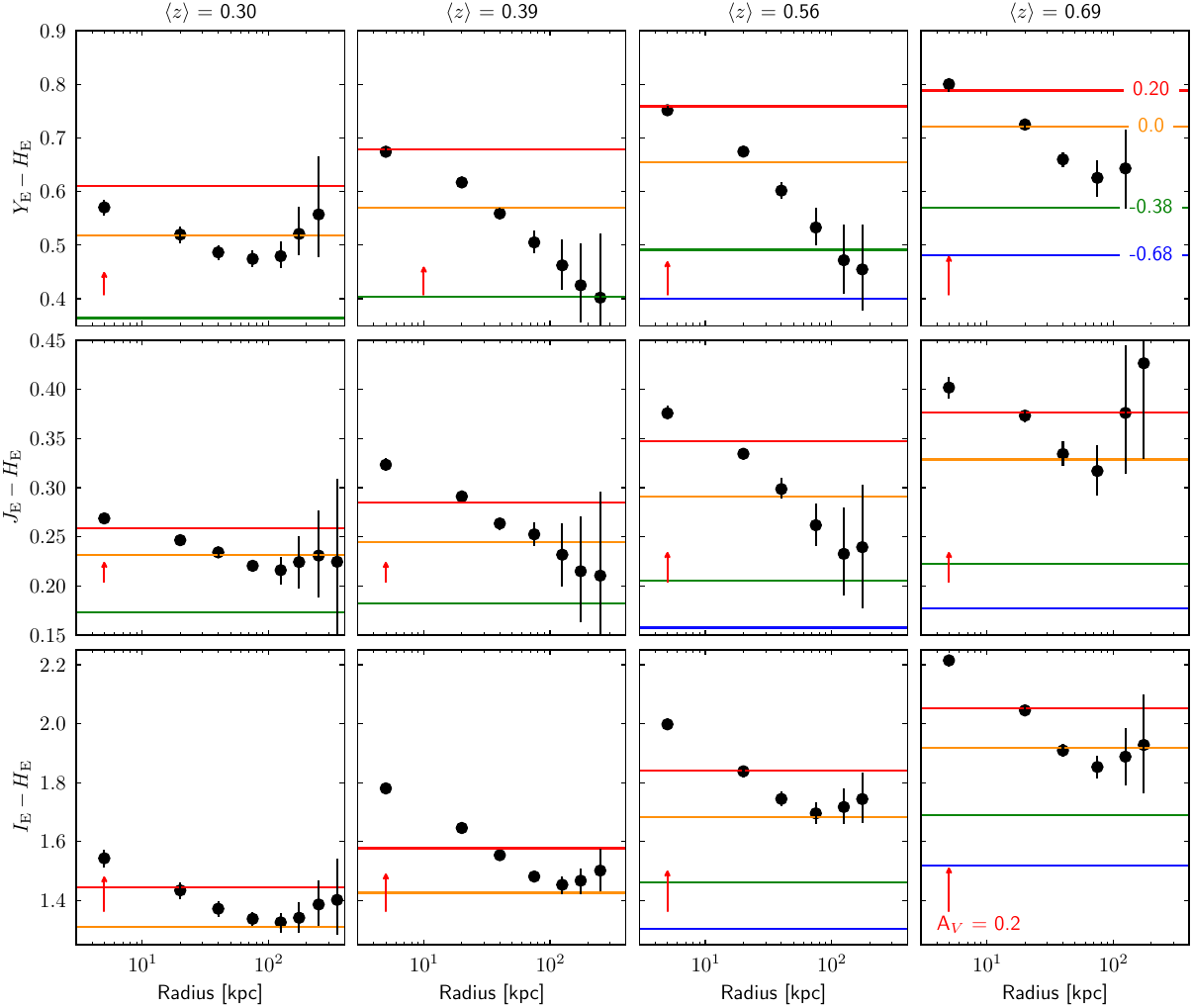}
    \caption{Stacked colour measurements as a function of the semi-major axis length of the annulus for four different redshift bins and three different colours. We include all points where 80\% of the bootstrap samples have both fluxes above S/N $>$ 10. The coloured lines in each row are the predicted colours (for a given redshift) for the E-MILES SPS models \citep{vazdekis2016} of fixed metallicity at $z_{\rm form} = 5$ based on the results shown in Fig.~\ref{fig:zform_Metallicity}. The values given on each line are the values of $\logten (Z/Z_\odot)$. The red arrow marks the impact on the colour of $A_{V} = 0.2\,\textrm{mag}$ of extragalactic dust extinction.}
    \label{fig:Colour_Gradient}
\end{figure*}

To avoid the bias that is caused by looking at individual clusters, and to extend our ICL colour measurements out to larger radii, we stacked our 1D light profiles in redshift bins. The flux in each stack was constructed from the mean of 42 clusters in the following redshift intervals: $0.1 < z < 0.348, 0.348 \le z < 0.432, 0.432 \le z < 0.63, 0.63 \le z < 0.75$. We did not apply any sigma clipping in our measurements. To measure the uncertainty in the average ICL colour, we performed 1000 bootstrap resamplings of our clusters and used the 16th and 84th percentile values as our uncertainties. To ensure that we were not biased by any particular cluster, we only retained radial colour data when more than 80\% of the bootstrap resampled measurements had fluxes in both bands that had a S/N $>$ 10. For the inner 150\,kpc where the weighted means and the stacks overlap, the two values provide consistent results. In Fig.~\ref{fig:Colour_Gradient}, we present the 1D stacked measurements and uncertainties on the average colour of the BCG+ICL for each redshift bin.

The colour of the average BCG+ICL profiles varies as a function of radius for each of the different redshift intervals. The location of each point on the $x$ axis represents the midpoint of the aperture. We can measure the colour of the ICL out to 400\,kpc at $z=0.3$, 300\,kpc at $z=0.39$ in the stacked profiles, but only out to 200\,kpc in the two higher redshift bins. 

Figure~\ref{fig:Colour_Gradient} presents a clear gradient in the average BCG+ICL colour between 10\,kpc and ${\approx}\,150$\,kpc in each of the three colours. At larger radii, the uncertainties are large enough that the data are consistent with the ICL colour either decreasing further or flattening out. In general, we note that the BCG (0--10\,kpc, 10--30\,kpc) is significantly redder than the outskirts, which are associated with the ICL. The presence of a radial gradient in the BCG+ICL colour is in strong agreement with prior analyses of both individual ICL studies and ICL populations, including stacked analyses \citep[e.g.][]{zibetti05, montes18, zhang18, demaio18, golden-marx23}. 

Since dust can redden the stellar colours, we also include a red arrow in Fig.~\ref{fig:Colour_Gradient}, which is representative of the impact of $A_{V} = 0.2\,\textrm{mag}$ of intrinsic extinction on each colour at each median redshift. As shown in \citet{gutierrez2017}, dust only contributes a small amount to the light within the ICL, as any dust in the ICM is expected to be sputtered by the X-rays from the ICM. In Fig.~\ref{fig:Colour_Gradient}, we can see that the impact of dust is quite small in the $\YE-\HE$ and $\JE-\HE$. However, the $\IE-\HE$ is much more susceptible to the impact of dust and it is possible that the colours are reddened as a result of dust.

\section{Relationship between NIR colours, metallicity, and age}
\label{sec:NIR}
Assuming that the dust content is very low, a colour gradient can be caused by either an underlying metallicity gradient, such that the stars that are bluer have a lower metallicity than the BCG, or the stars in the outskirts are younger than the old `red and dead' stellar population in the core of the BCG, or a combination of the two, as shown in \citet{montes18}. Using the \texttt{E-MILES} stellar population synthesis (SPS) models \citep{vazdekis2016}, which assume a Kroupa universal initial mass function \citep{kroupa2001}, we estimated how the colour of the stellar population depends on both formation redshift, $z_{\rm form}$, and the metallicity. In this model we assumed a single burst of star formation that occurs at $z_{\rm form}$. We note that it is possible that a more complex star-formation model is required to fully model the colour of the ICL. However, the inclusion of additional parameters, such as bursts of star formation, would likely lead to degeneracies between potential models. We calculated each model for a specific observed redshift. We note that these models do not account for specific abundance element ratios and assume that [M/H] = [Fe/H] \citep{vazdekis2016}. Thus, we only constrained the total metallicity, not the amount of Fe. 

\begin{figure*}
    \centering
    \includegraphics[width=18cm,trim={0cm 0 0cm 0}]{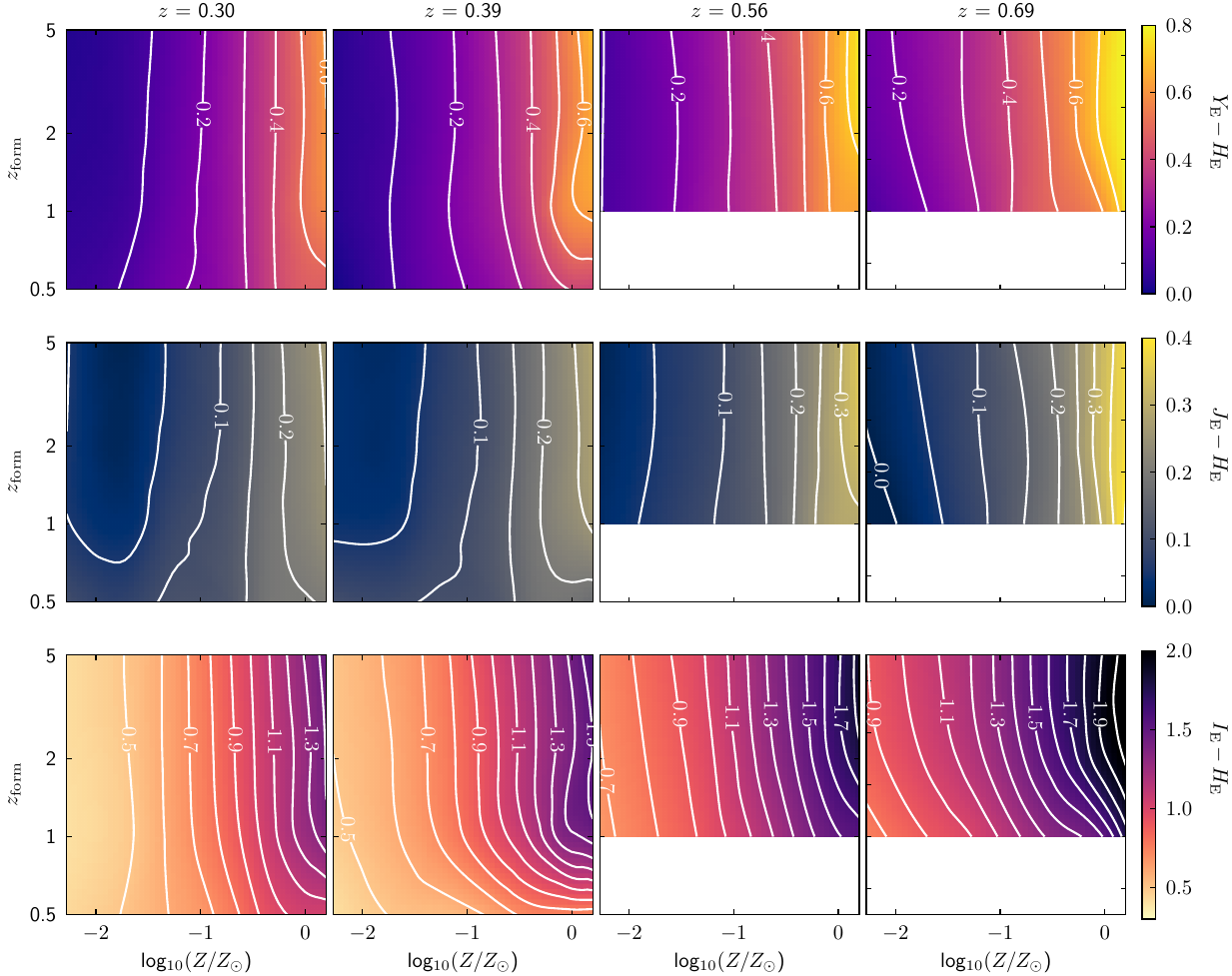}
    \caption{Metallicity, $\logten(Z/Z_\odot)$, $z_{\rm form}$, and photometric colour (as given by the colour bar for each row) parameter space. We show four different redshift bins, based on those of our observations, as well as the three different colours used in this analysis. We used the E-MILES SPS models \citep{vazdekis2016} to construct this parameter space. The white lines illustrate contours of fixed colour to highlight how photometric colour depends on $z_{\rm form}$ and metallicity. For the last two columns, we do not include where $z_{\rm form} < 1$ because this would result in very young stellar populations that are not characteristic of the BCG or ICL.}
    \label{fig:zform_Metallicity}
\end{figure*}

In this analysis, using the four \Euclid bands we chose to measure the $\YE-\HE$, $\JE-\HE$, and $\IE-\HE$ colours. We constructed SEDs at observed redshifts of $z=0.3$, $0.39$, $0.56$, and $0.69$ using the E-MILES SPS models \citep{vazdekis2016}. The models have fixed stellar metallicities of $\logten(Z/Z_\odot) = 0.2$, $0.0$, $-0.38$, $-0.68$, $-1.28$, $-1.68$, and $-2.28$, and the stellar ages were fixed such that the stars formed at a redshift of 0.5, 1, 2, 5, or 10. We determined the three colours by convolving the SEDs with the filter transmission curves of the \Euclid filters. Finally, we performed a smooth interpolation across the metallicity-formation redshift space to determine how the stellar colours depend on the parameter space at the median redshift of the four bins used in this analysis. The results are shown in Fig.~\ref{fig:zform_Metallicity}. While the colour bars in Fig.~\ref{fig:zform_Metallicity} show the photometric colour of a given stellar population model, we also overlay colour isocontours (in white), to show how a fixed photometric colour depends on both $z_{\rm form}$ and metallicity. 

For the three \Euclid colours we explore, we find that there is a strong relationship between metallicity and colour, with the higher-metallicity stars having redder colours. However, there is no coherent relationship between the colours and stellar formation redshift. For $\YE-\HE$ and $\JE-\HE$, the colour does not depend on $z_{\rm form}$ if the stellar population formed at $z_{\rm form} > 1$ \citep[e.g.][]{Anders04}. This can be seen as the colour contours are almost vertical lines in the metallicity-$z_{\rm form}$ space. When the stellar population formed at $z_{\rm form} < 1$, there is a relationship between colour and stellar age: the higher-metallicity stars tend to be bluer at younger stellar ages, whereas for the lowest-metallicity stars, the younger stars tend to be slightly redder in $\JE-\HE$, but have no dependency on age for the $\YE-\HE$ colour. At $z < 0.39$, the $\IE-\HE$ colour is mostly independent of stellar age when the stars formed at $z_{\rm form} > 1$, but at $z_{\rm form} < 1$ the colour depends strongly on stellar formation redshift, with younger stars having bluer colours for all metallicities. At $z\ge0.56$ the $\IE-\HE$ colour depends both on stellar age and metallicity unless the stars have a formation redshift of $z_{\rm form} \gtrsim 3$, which can be seen as the colour contours begin to tilt away from a vertical line. 

In summary, if the formation redshift of the stellar population is greater than 1, then the $\YE-\HE$ and $\JE-\HE$ colours can constrain the metallicity of the ICL. On the other hand, the $\IE-\HE$ colour may be able to provide some weak constraints on the stellar age. 

\section{Age and metallicity of the ICL}
\label{sec:Discussion}

The $\IE-\HE$ colours presented in Fig.~\ref{fig:Colour_Gradient} are very red, even for the ICL between 200\,kpc and 400\,kpc. By comparing these measured colours to the isocontours in the bottom panel of Fig.~\ref{fig:zform_Metallicity}, we see that these observed colours imply that the ICL stars were formed at a redshift of $z_{\rm form} > 1$. Dust extinction would impact this colour (see red arrow in Fig.~\ref{fig:Colour_Gradient}), but we do not expect dust to be significant in the ICL as X-rays from the ICM would sputter the dust grains. Therefore, the $\IE-\HE$ colour implies that the stellar population of the ICL is relatively old, with a formation epoch of $z_{\rm form} > 1$. This finding is supported by simulations and observations, which find that the median age of the ICL is of the order of 10 Gyr \citep[e.g.][]{edwards20, brown2024}. Given that we have constrained the stellar population to have a formation epoch of $z_{\rm form} > 1$, we are now able to use the $\YE-\HE$ and $\JE-\HE$ colours to constrain the metallicity of the ICL (Fig.~\ref{fig:zform_Metallicity}). In Fig.~\ref{fig:Colour_Gradient} we mark the expected colour corresponding to a range of metallicities, assuming the median redshift of each bin and that all the stars formed at $z_{\rm form} = 5$. Our conclusions do not change if we assume $z_{\rm form}$ to be in the range of $2 < z_{\rm form} < 10$. 

\begin{figure*}
    \centering
    \includegraphics[width=18cm, trim={0cm 0 0cm 0}]{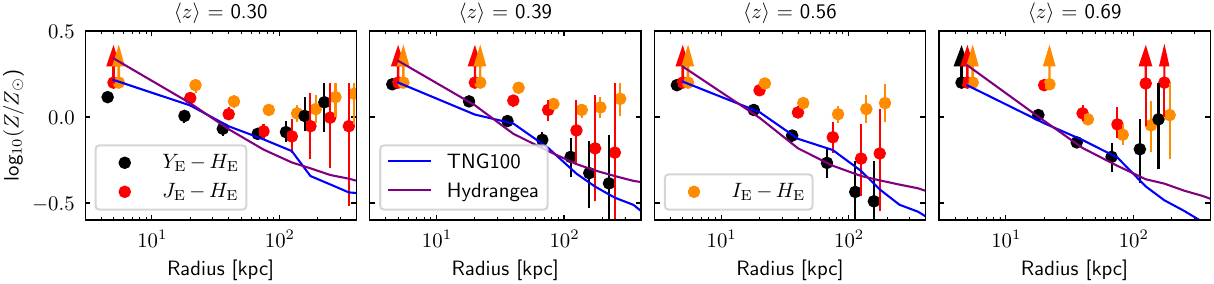}
    \caption{Metallicity of the ICL as a function of the semi-major axis length of the annuli for four different redshift bins. The red points represent the $\JE-\HE$ measurements, the black points are the $\YE-\HE$ measurements, and the orange points represent the $\IE-\HE$ measurements. The blue and purple lines represent the mean luminosity-weighted metallicity profile of clusters predicted by the TNG100 simulation and Hydrangea simulation at the respective redshift. We apply a slight offset in the $x$ position to the data to highlight the differences between the colour measurements.}
    \label{fig:metallicity_gradient}
\end{figure*}

We use the values of $\YE-\HE$, $\JE-\HE$ and $\IE-\HE$ measured in each physical aperture and read off the metallicity that each colour corresponds to within the colour-metallicity space shown in Fig.~\ref{fig:zform_Metallicity}, along the line of $z_{\rm form} = 5$. We note that as a result of our choice of SPS models, we have an upper limit on the metallicity of $\logten (Z/Z_\odot) = 0.20$. Thus, for some of the inner regions, we are unable to adequately constrain the metallicity and can only place a lower limit on the value (as shown by the red arrows in Fig.~\ref{fig:metallicity_gradient}).

In Fig.~\ref{fig:metallicity_gradient}, we present the metallicity of the BCG+ICL based on the colour measurements of the 1D stacked light profiles. The error bars on the metallicity are based on the bootstrap uncertainty shown in Fig.~\ref{fig:Colour_Gradient}. We measure a metallicity profile out to ${\approx}\,300$\,kpc for the first time using a statistical sample and see a metallicity gradient in all three colours.
We note that there is some variation between the three colours and across the different redshift ranges. Generally, the $\IE-\HE$ and $\JE-\HE$ colours yield slightly higher metallicities. However, across all three colours, we detect a metallicity gradient between 10\,kpc and ${\approx}\,150$\,kpc such that the core of the BCG is super-solar and the ICL (beyond 100\,kpc) is sub-solar. At larger radii, we are unable to fully determine the trends as either a flattening in metallicity gradient or continuously decreasing metallicity is consistent within the large uncertainties. Despite this uncertainty, we conclude that using \Euclid colours has allowed us to put a lower limit on the stellar age of the ICL, and detect a radial gradient in metallicity.

While the detection of a metallicity gradient is independent of our choice of SPS model, the metallicity values estimated from these models and presented here are dependent on our choice of SPS model and colour. In Appendix~\ref{app:SPS}, we show the metallicity measurements when assuming a \citet{bruzual03} model, which would yield a similar gradient, but differing metallicities of the order of 0.2\,dex. Thus, we emphasise that while this metallicity gradient is a real feature of the BCG+ICL, including more parameters in our SPS models (such as multiple bursts of star formation) and additional optical photometry bands is needed to improve estimation of the metallicity measurements. Furthermore, it is possible that spectroscopic measurements of the ICL may be needed to break the degeneracies that exist between more complex star-formation histories. 

In Fig.~\ref{fig:metallicity_gradient}, we overlay the 2D projected luminosity-weighted mean metallicity profiles for the BCG+ICL in the TNG100 simulation of the \textsc{IllustrisTNG} Project \citep[in blue;][]{nelson18, pillepich17, springel18, Nelson2018a, naiman2018, marinacci2018} and the Hydrangea simulation \citep[in purple;][]{bahe17}. Hydrangea is a zoom-in simulation of galaxy clusters that have been evolved within the EAGLE galaxy formation model \citep{schaye2015}. For both simulations, the measurements are based on a single 2D projection for each cluster. To separate ICL particles from those of the BCG and galaxies, for TNG, we followed \citet{Brown2026} and applied a threshold in instantaneous local stellar density to identify ICL emulating \texttt{AdaptaHOP} \citep{Dubois2014}, while for Hydrangea, we used an unbinding threshold based on the instantaneous kinetic and potential energies. Though these are differing ICL definitions, both simulations show a clear gradient between the metallicity of the BCG core and that of the ICL, in agreement with our observations. For the TNG100 measurement, we measure the mean luminosity weighted metallicity based on the 15 most massive halos [at $z=0$ the halo mass range is $13.95 < \logten(M_{\rm halo}/M_\odot) < 14.6$]. For Hydrangea, we measure the mean luminosity weighted metallicity based on 54 clusters with $\logten(M_{\rm halo}/M_\odot) > 14.0$. We see that the BCG (inner 10\,kpc) metallicity is slightly higher in Hydrangea than TNG; however, we note that the metallicities from Hydrangea can be biased high \citep{schaye2015}, which may result in this difference.  

\begin{figure}
    \centering
    \includegraphics{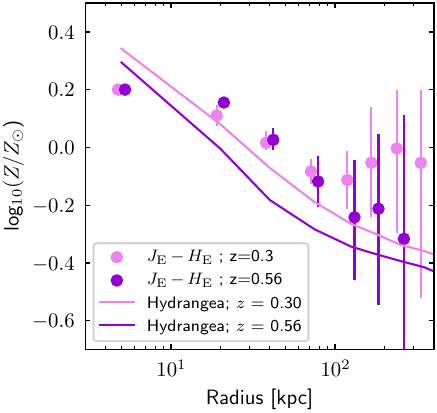}
    \caption{Metallicity, $\logten(Z/Z_\odot)$, of the ICL as a function of the semi-major axis length for our lowest and second-highest redshift bins. The lines represent the values from the Hydrangea simulation and the points are the values measured from \Euclid for the $\JE-\HE$ measurements. The darker colour is for the higher redshift measurements, while the lighter colour is the lower redshift measurements. We apply a slight offset in the $x$ position to emphasise the difference between the redshift bins.}
    \label{fig:metallicity_redshift}
\end{figure}

In Sect.~\ref{sec:intro}, we noted a few predictions about the metallicity of the ICL from simulations and models. (i) We similarly find that the ICL (the region beyond 100\,kpc) is ${\approx}\,0.4$\,dex more metal-poor than the BCG (inner 30\,kpc). (ii) We detect a persistent metallicity gradient out to ${\approx}\,150$\,kpc, in agreement with what has been observed in TNG and Hydrangea, as shown in Fig.~\ref{fig:metallicity_gradient}. However, we note that beyond 150\,kpc, we are unable to conclude that a gradient is present. (iii) In Fig.~\ref{fig:metallicity_redshift}, we compare the metallicity gradient for the simulations and the $\JE-\HE$ measurements from \Euclid for the lowest and second-highest redshift bins\footnote{Since the ICL is brighter in the $\JE$ band than the $\YE$ band, and because the potential impact of dust on the $\JE-\HE$ is smaller than the $\YE-\HE$ (see Fig.~\ref{fig:Colour_Gradient}), we focus only on this colour for our further analyses.}. There is a slight evolution in the metallicity gradient in the simulations, such that the ICL is more metal-poor at higher redshifts. The observations agree with this trend, but there is not sufficient signal to confirm it exists in the Universe. Since this comparison suggests the possibility of evolution, we note that dividing the radius by $R_{\rm 200m}$ removes much of this offset in the simulations and decreases it in the observations. As the offset does not entirely disappear in observations, this suggests the evolution of the metallicity gradient is not entirely self-similar. However, our metallicity measurements within each redshift bin at large radii are within $1\sigma$ of one another, so it remains unclear whether these results are statistically robust. Thus, using the upcoming DR1 data (1900\,deg$^2$) and cluster catalogues will significantly increase the sample size, allowing us to investigate whether this change in the metallicity gradient is real. (iv) Our results show a steep (non-zero) metallicity gradient, indicating that the ICL is likely built predominantly via tidal stripping.  

\subsection{Comparison to previous observational results}
As noted in Sect.~\ref{sec:intro}, a colour gradient within the ICL has been commonly detected in both individual observations and large statistical studies, such that the ICL is bluer than the core of the BCG \citep[e.g.][]{zibetti05, montes18, demaio18, zhang19, ragusa21, golden-marx23}. As a result of our stacking and tight constraints on the uncertainty of the flux measurements at large radii, we measure a colour profile out to ${\approx}\,300$\,kpc and detect a colour gradient out to ${\approx}\,150$\,kpc. Beyond these radii, we are unable to adequately determine whether the gradient in colour and metallicity persists or flattens.

Before \Euclid it was challenging to characterise the metallicity within the ICL because of the degeneracy that exists between the age and metallicity of the ICL when estimated using optical colours \citep[e.g.][]{contini2021}. Therefore, there are only a few single target or small sample studies, most of which only extend out to ${\approx}\,200$\,kpc, with which we can directly compare our results. Specifically, \citet{Montes2014} and \citet{montes18} detect both a metallicity and an age gradient in individual ICL measurements using the HST Frontier Fields clusters, which are generally more massive than those observed in this analysis, while \citet{demaio15} assume that the age is fixed when detecting a metallicity gradient in their small ICL sample. Moreover, while ICL metallicity measurements are not commonplace, there is a strong consensus among spectroscopic studies that the BCG is more metal-rich (super-solar) compared to the outskirts of the ICL, which is sub-solar \citep[e.g.][]{coccato2010, edwards2016, edwards20, edwards24, Johnston2018, gu20}.

\begin{figure*}
    \centering
    \includegraphics[width=18.0cm]{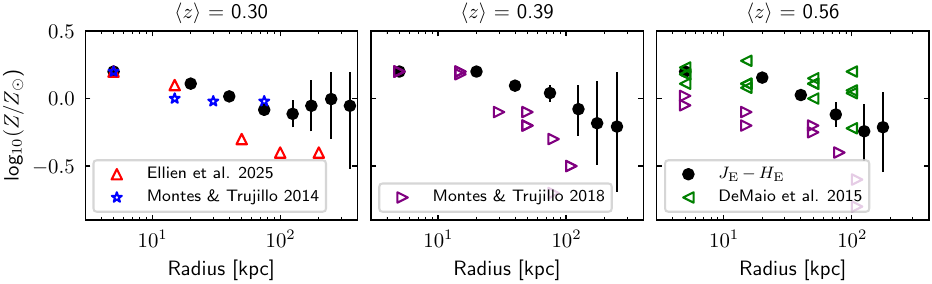}
    \caption{Metallicity, $\logten(Z/Z_\odot)$, of the ICL as a function of the semi-major axis length for our three lowest redshift bins. The black points are the values measured from \Euclid for the $\JE-\HE$ measurements. The coloured points are the directly comparable values from the literature.}
    \label{fig:metallicity_comp}
\end{figure*}

In Fig.~\ref{fig:metallicity_comp}, we compare the metallicities derived from our $\JE-\HE$ colours to the measurements from the literature that cover a similar redshift range \citep{Montes2014, montes18, demaio15, ellien25}. This figure shows that our measured BCG and ICL metallicities, as well as the presence of a metallicity gradient, are in good agreement with these previous observational studies for the region of overlap. 
Moreover, our results are similar to those from \citet{EROPerseusICL}, who use the $\JE-\HE$ and $\YE-\JE$ colours and estimate the metallicity of the ICL in Perseus as $\logten (Z/Z_\odot) \approx -0.8$ at ${\approx}\,150$\,kpc. Additionally, while our results agree with prior measurements, they do highlight that in order to fully characterise the metallicity profile of the ICL, large samples of clusters must be used.

Building on this analysis, there are a number of questions we plan to explore to improve our understanding of ICL formation with the upcoming data from Euclid DR1. In particular, the larger and more statistically robust dataset from DR1 will extend these stacked ICL measurements out to much larger radii. These added data should also allow us to address whether the ICL colour and metallicity are influenced by either the cluster mass as suggested by some models, simulations, and observations \citep{contini14, contini19, Cui14, ragusa23} or the dynamical state of the cluster, estimated using parameters including the magnitude gap or ICL fraction \citep[e.g.][]{Golden-Marx25, kimmig25}, further enhancing our understanding of how the underlying dark matter halo correlates with and influences the formation of the ICL.  

\subsection{Formation history of the ICL}
The current consensus, based on observations and simulations, is that the ICL grows as a result of tidal stripping \citep[e.g.][]{contini19, brown2024, Contreras-Santos2025}. Moreover, simulations suggest that the majority of ICL stars come from massive [$\logten (M_\star/M_\odot) > 10.5$] satellite galaxies, though we note that the fractional mass loss from lower-mass galaxies is higher \citep[e.g.][]{contini14, contini19, brown2024, Brown2026, Contreras-Santos2025}. Based on the colour gradient and sub-solar metallicity, our results strongly support the idea that the tidal stripping of satellite galaxies is the primary growth mechanism for the ICL out to ${\approx}\,150$\,kpc. Using our ICL metallicities in the 200--300\,kpc region, we estimated the progenitor galaxy population using the mass-metallicity relation from \citet{Gallazzi2005}, which is based on observations of the local Universe, to be $9.7 < \logten (M_\star/M_\odot) < 10.5$. However, since the progenitor satellite galaxies are unlikely to be entirely stripped, just the metal-poor outskirts, these stellar mass estimates are lower limits, which thus align with simulations. 

In this analysis, we have placed constraints on the metallicity of the ICL out to large radii. However, the NIR and limited available optical data do not allow us to strongly constrain the age of the stellar population. Thus, we present an explanation for the physical formation of the metallicity gradient that is not dependent on age.
If, based on the presence of the colour and metallicity gradient between 10\,kpc and ${\approx}\,150$\,kpc, we assume that at least part of the ICL forms predominantly via tidal stripping, then the outskirts of these satellite galaxies will be stripped upon infall into the cluster, because they lie in a shallower part of the gravitational potential well of the cluster \citep{martin24}, and those stars will populate the ICL. Meanwhile, the more metal-rich satellite galaxy cores lose energy through dynamical friction \citep[e.g.][]{contini2021} and either merge with the BCG \citep{brown2024} or, depending on their orbits and infall velocities, remain as satellites.
Following this model, there are two unique physical processes that are responsible for growing the BCG+ICL: the inner region assembles as a result of the hierarchical merging of the cores of satellite galaxies with the BCG (yielding a metal-enhanced core), while the outer region (the ICL) grows via the tidal stripping of the outskirts of satellite galaxies, in agreement with previous results \citep{Golden-Marx25}. However, we note that deeper data from Euclid DR1 will inform whether the colour and metallicity measurements beyond ${\approx}\,150$\,kpc are consistent with this formation picture. 

\section{Conclusions} \label{sec:conclusion}
In this work, we use the \Euclid Q1 observations of 168 DES-redMaPPer clusters over the redshift range $0.1 < z < 0.75$ with $\logten (M_{\rm halo}/M_\odot/h) > 14.05$ to determine how the stellar properties (colour, age, and metallicity) of the ICL vary with radius and redshift to inform the progenitor population that forms the ICL. We summarise the primary results of this analysis as follows. 
\begin{itemize}
    \item Using individual light profiles, robust ICL colours can be measured for 75\% of the clusters out to 100\,kpc, but this rapidly decreases to ${<}\,10$\% at 200\,kpc. We find that the ICL signal is predominantly detectable among both more massive and lower-redshift clusters at larger radii.
    \item Using the individual profiles, we see that the colours of the inner 30\,kpc, which is representative of the BCG, are remarkably consistent. Beyond this extent, the colours have a large scatter. This suggests that the stellar population of the ICL has large cluster-to-cluster variation, and therefore care should be taken not to infer average ICL properties from small samples or individual clusters.
    \item Using stacked 1D light profiles allows us to robustly measure the ICL colours out to ${\approx}\,300$\,kpc (Fig.~\ref{fig:Colour_Gradient}) and detect a colour gradient in each of the $\YE-\HE$, $\JE-\HE$, and $\IE-\HE$ colours between 10\,kpc and ${\approx}\,150$\,kpc. At larger radii, we are unable to consistently characterise the colour trend as the uncertainties mean the profiles are consistent with either a flattening in the trend or a continually decreasing colour with radius.
    \item The red $\IE-\HE$ colour of the outer ICL (Fig.~\ref{fig:Colour_Gradient}) implies the ICL stellar population must have formed at a redshift earlier than $z=1$, based on the E-MILES SPS models \citep{vazdekis2016}.
    \item Using the E-MILES SPS models \citep{vazdekis2016}, we show that at $z_{\rm form} > 1$ a NIR colour gradient can be explained by a metallicity gradient, which suggests that this trend may be independent of the age of the stellar population, potentially breaking the degeneracy that exists between colour, age, and metallicity.
    \item Using the metallicity--$z_{\rm form}$--colour parameter space (Fig.~\ref{fig:zform_Metallicity}), we determined how the observed colour profiles (see Fig.~\ref{fig:Colour_Gradient}), assuming the E-MILES SPS models \citep{vazdekis2016}, transform into a metallicity profile (see Fig.~\ref{fig:metallicity_gradient}). We find that the core of the BCG has a super-solar metallicity, while the outskirts of the ICL are metal-poor. Using this stacked data, we measured how the metallicity of the BCG+ICL changes as a function of radius within four redshift bins. In each of the three colours, we detect a clear metallicity gradient out to ${\approx}\,150$\,kpc. We place the first statistical constraints on the metallicity of the ICL population out to ${\approx}\,300$\,kpc from the BCG. These metallicity constraints can be used to constrain galaxy formation and merger history models of the ICL and also improve ICL stellar mass estimates in future observations.  
    \item Based on the metallicity of the ICL and the colour and metallicity gradients, our results support the idea that the ICL (at least out to ${\approx}\,150$\,kpc) grows through the tidal stripping of satellite galaxies while the BCG (as supported by its enhanced metallicity) grows via mergers with the dense metal-rich cores of satellite galaxies. 
\end{itemize}
We find that the NIR colours from \Euclid statistically constrain the metallicity of the ICL. To constrain the age along with the metallicity, we need to sample more of the spectral energy distribution of the ICL. By combining the \Euclid NIR colours with the deeper optical colours -- using narrower filters -- from the Vera C. Rubin observatory \citep{LSST}, we expect to be able to accurately determine the ages of the stellar population within the ICL, which will give us all the information needed to fully characterise the stellar population within the ICL.  

\begin{acknowledgements}

The authors would like to thank the anonymous referee for their feedback on this paper, which helped to greatly strengthen the presented analysis.
J.B.G.M., N.A.H., T.K., H. Gao, H. Gully, M. U., and M.B. gratefully acknowledge financial support from The Leverhulme Trust. 
M.M. acknowledges support from grant RYC2022-036949-I financed by the MICIU/AEI/10.13039/501100011033 and by ESF+, grant CNS2024-154592 financed by MICIU/AEI/10.13039/501100011033, and program Unidad de Excelencia Mar\'{i}a de Maeztu CEX2020-001058-M.  

Y.M.B. acknowledges support from UK Research and Innovation through a Future Leaders Fellowship (grant agreement MR/X035166/1) and financial support from the Swiss National Science Foundation (SNSF) under project 200021\_213076. This work used the DiRAC@Durham facility managed by the Institute for Computational Cosmology on behalf of the STFC DiRAC HPC Facility (www.dirac.ac.uk). The equipment was funded by BEIS capital funding via STFC capital grants ST/K00042X/1, ST/P002293/1, ST/R002371/1 and ST/S002502/1, Durham University and STFC operations grant ST/R000832/1. DiRAC is part of the National e-Infrastructure.  

W.C. gratefully acknowledges Comunidad de Madrid for the Atracci\'{o}n de Talento fellowship no. 2020-T1/TIC19882 and Agencia Estatal de Investigaci\'{o}n (AEI) for the Consolidaci\'{o}n Investigadora Grant CNS2024-154838. He further acknowledges the Project PID2024-156100NB-C21 financed by MICIU/AEI /10.13039/501100011033/FEDER, EU and ERC: HORIZON-TMA-MSCA-SE for supporting the LACEGAL-III (Latin American Chinese European Galaxy Formation Network) project with grant number 101086388 and the science research grants from the China Manned Space Project.

F.D. acknowledges support from CNES.

A.H.G. acknowledges the support of NASA ROSES Grant 12-EUCLID11-0004.

L.M. acknowledges the financial contribution from the PRIN-MUR 2022 20227RNLY3 grant “The concordance cosmological model: stress-tests with galaxy clusters” supported by Next Generation EU and from the grant ASI n. 2024-10-HH.0 “Attività scientifiche per la missione \Euclid - fase E”.

R.R. acknowledges financial support through grants PRIN-MIUR 2020SKSTHZ and through INAF WEAVE StePS funds.

\AckQone
\AckEC

The IllustrisTNG simulations were undertaken with compute time awarded by the Gauss Centre for Supercomputing (GCS) under GCS Large-Scale Projects GCS-ILLU and GCS-DWAR on the GCS share of the supercomputer Hazel Hen at the High Performance Computing Center Stuttgart (HLRS), as well as on the machines of the Max Planck Computing and Data Facility (MPCDF) in Garching, Germany.

\end{acknowledgements}

\bibliography{aa60166-26}

\newpage
\appendix 
\onecolumn

\section{Impact of SPS model selection}
\label{app:SPS}
\begin{figure*}[t]
    \centering
    \includegraphics[width=\textwidth,trim={0cm 0 0cm 0}]{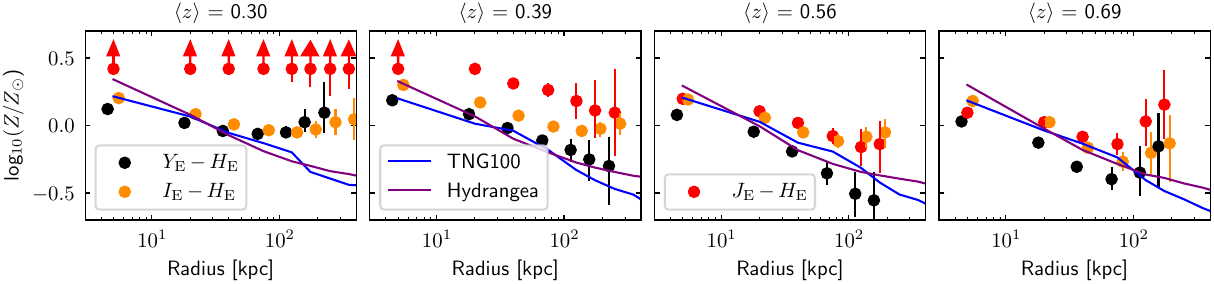}
    \caption{Metallicity of the ICL as a function of the semi-major axis length of the annuli for four different redshift bins when estimated using the \citet{bruzual03} SPS model. The red points represent the $\JE-\HE$ data, the black points are the $\YE-\HE$ measurements, and the orange points represent the $\IE-\HE$ data. The blue and purple lines represent the mean luminosity-weighted metallicity profile of clusters predicted by the TNG100 simulation and Hydrangea simulation at the respective redshift. Red arrows are used for data that are above our maximum metallicity value. We also add a slight offset to the $x$-position of the data to distinguish between the metallicities. }
    \label{fig:IH_met_BC03}
\end{figure*}
We investigate the impact of alternative choices of SPS models on our estimate of the metallicity of the stellar population of the BCG and ICL. In the main text of this analysis, we used the E-MILES SPS models \citep{vazdekis2016}. Here, we use a \citet{bruzual03} SPS model. We note that we follow the same methodology as discussed in Sects.~\ref{sec:NIR} and ~\ref{sec:Discussion}. In Fig.~\ref{fig:IH_met_BC03}, we present the metallicity distributions for the same three colours as in Fig.~\ref{fig:metallicity_gradient}, but estimated using a \citet{bruzual03} SPS model. We note that the maximum metallicity for this model is ${\approx}\,0.4$, which is greater than that of the E-MILES SPS models \citep{vazdekis2016}. As shown, we see that the metallicities based on the $\IE-\HE$ and $\YE-\HE$ are more consistent at lower redshifts than the $\JE-\HE$ values. This is a result of degeneracies that exist within the \citet{bruzual03} models for NIR wavelengths since that portion of the SED is based on extensions between the STELIB and Pickles spectral libraries using the BaSeL library and the higher wavelength Pickles library is of only one metallicity. These same degeneracies do not exist within the E-MILES models, which are derived from the Infrared Telescope Facility Spectral Library \citep{vazdekis2016}. We note that at higher redshifts, the NIR colours are redshifted to lower wavelengths that are not in the regions covered by these extensions. However, regardless of the difficulty of using the $\JE-\HE$ colours at lower redshifts, we find nearly identical trends, in terms of a metallicity gradient, its extent, and trends beyond ${\approx}\,150$\,kpc to what we detect when using the E-MILES SPS models \citep{vazdekis2016}. As such we find that our detection of a metallicity gradient is independent of our choice of SPS model. However, we note that the actual metallicity values are dependent on the choice of SPS model and differ by ${\approx}\,0.2$\,dex (when not considering the $\JE-\HE$ colours).

\label{lastpage}
\end{document}